\documentclass[granma]{svjour}
\usepackage{latexsym,amssymb,amsmath}
\usepackage{graphics,epic,eepic}
\begin{document}
\title{Axial Particle Diffusion in Rotating Cylinders}
\author{Christian M. Dury and Gerald H. Ristow
}                     
%
%
\institute{Fachbereich Physik \\ Philipps-Universit\"at \\ 
           35032 Marburg, Germany}
\date{Received: \today / Revised version: date}
%
\maketitle
\begin{abstract}
We study the interface dynamics of a binary particle mixture in a rotating
cylinder numerically. By considering only the particle motion in axial
direction, it is shown that the initial dynamics can be well described by a
one-dimensional diffusion process. This allows us to calculate a {\em
macroscopic}\/ diffusion constant and we study its dependence on the
inter-particle friction coefficient, the rotation speed of the cylinder and the
density ratio of the two components. It is found that radial segregation
reduces the drift velocity of the interface. We then perform a {\em
microscopic}\/ calculation of the diffusion coefficient and investigate its
dependence on the position along the cylinder axis and the density ratio of the
two particle components. The latter dependence can be explained by looking at 
the different hydrostatic pressures of the two particle components at the
interface. We find that the microscopically calculated diffusion
coefficient agrees well with the value from the macroscopic definition when
taken in the middle of the cylinder.
\end{abstract}
\section{Introduction}
\label{sec: intro}
A common device used for mixing different kinds of materials is a rotating kiln
or cylinder~\cite{bridgwater76} where the mixing rate and the particle dynamics
depend on the rotation speed of the cylinder~\cite{rose59,nityanand86}.
However, when materials which differ in size or density are used, particles
with different properties tend to accumulate in different spatial regions which
is called {\em segregation}. Two different types of segregation are commonly
observed in rotating cylinders:
\begin{enumerate}
  \item[(a)] a fast {\bf radial} segregation
  \item[(b)] a much slower {\bf axial} segregation.
\end{enumerate}
The latter leads to band formation along the rotational axis and it takes many
cylinder rotations before a steady state with respect to the observed band
structure is  reached
\cite{oyama39,donald62,dasgupta91,nakagawa94,hill95,fabi97,choo97}. In the case
of radial segregation, it usually takes only a few rotations to reach a fully
segregated state where the smaller or denser particles form a central core
right below the fluidized surface layer. This was studied experimentally and
numerically for varying size
ratios~\cite{clement95,cantelaube95,baumann95,hill97,dury97,dury98} and density
ratios~\cite{ristow94,metcalfe96,khakhar97b}. The amount and direction of
segregation depends on the rotation rate~\cite{nityanand86}.

The common method to study the segregation process starts from a well mixed
state and records the segregation amount or the spatial pattern as function of
time. This works well in the case of {\em radial} segregation and quantitative
results regarding the dependence of the segregation process on rotation speed
and size ratio were obtained using a suitable, normalizable order parameter
$q_\infty$~\cite{dury97,dury98}. However, the {\em axial} segregation process
is much richer due to the three-dimensional particle motion and small changes
in the initial mixture seem to have a large effect on the band formation
process. The final number of bands, their positions and widths varied from
experiment to experiment. For example Nakagawa~\cite{nakagawa94} found that a
three  band configuration was the most stable after an extended number of
rotations.  These bands are normally not pure and a radially segregated core of
smaller or/and denser particles might still be present~\cite{hill97}. Chicharro
et al.~\cite{fabi97} rotated two sizes of Ottawa sand for two weeks at 45
revolutions per minute (rpm) and found a final state of two {\em pure} bands
each filling approximately half of the cylinder, i.e.\ {\em no} radial core was
found.

Depending on the particle kinds used in the experiments, the band formation
process is more or less pronounced and for some combinations not observed at
all. Different explanations have been proposed: 
\begin{enumerate}
\item Donald and Roseman~\cite{donald62} concluded from their experiments that 
      no banding occurs when the smaller particles have a smaller {\em static 
      angle of repose}; 
\item das Gupta et al.~\cite{dasgupta91} modified this statement by saying that 
      the relevant quantity is the difference in surface angle of the two 
      components at a specific rotation speed ({\em dynamic angle of repose}) 
      and 
\item Hill and Kakalios~\cite{hill95} proposed a model based on a diffusion 
      equation with an {\em effective diffusion coefficient} to account for 
      their finding of ``reversible axial segregation''. 
\end{enumerate}
Recently it was argued that other transport mechanisms can drive the
segregation process, especially that avalanches play a role~\cite{frette97}.

\begin{figure}
  \resizebox{\hsize}{!}{%
    \includegraphics{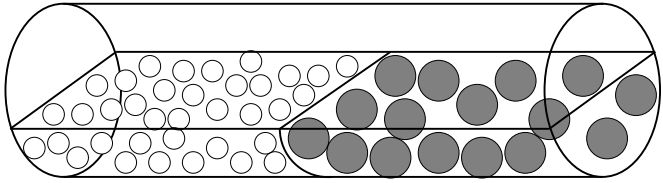}
  }
  \caption{Sketch of the initial configuration: large particles are all in the 
           right half of the cylinder and shown in gray.}
  \label{fig: sketch}
\end{figure}

In order to have a better defined initial configuration for binary mixtures, it
was proposed to fill one half of the cylinder with one particle component and
the other half with the other component~\cite{nakagawa97} which is sketched in
Fig.~\ref{fig: sketch} for a system with particles of different sizes. Such a
configuration allows for a detailed analysis of the individual dependencies and
will be used as initial state throughout this paper to study the mixing and
segregation of binary particle mixtures along the rotational axis numerically.

The paper is organized in the following way: In the next section, we  describe
our numerical model which solves Newton's equation of motion for each particle
and discuss the physical interpretation of our simulation parameters. In
section~\ref{sec: angle}, we demonstrate how different dynamic angles of
repose  can be obtained using our numerical model and compare their values to
experiments. In section~\ref{sec: mixing}, the particle dynamics along the
rotational axis are described by a one-dimensional diffusion equation and the
calculated diffusion constant are studied as function of the inter-particle
friction coefficient, the rotation speed of the cylinder and the density ratio
of the two components. It is also compared in section~\ref{sec: macro} to a
microscopic calculation of the diffusion coefficient based on the individual
particle motion and the agreement is very satisfactory. The conclusions round
off the paper. 

\section{Numerical Model}
\label{sec: model}
We use three-dimensional discrete element methods, also known as {\em granular
dynamics}~\cite{ristow94b}, which gives us the advantage to vary particle
properties like density and friction coefficient freely, whereas in experiments
the number of different kinds of beads is rather limited.

Each particle $i$ is approximated by a sphere with radius $R_i$. Only  contact
forces during collisions are considered and the particles are allowed to
rotate; we also include rolling resistance to our model to correctly describe
the particle rotations (see Ref.~\cite{dury98,dury98b}). The forces acting on
particle $i$ during a collision with particle $j$ are 
\begin{equation}
  F_{ij}^n = - \tilde{Y} \ (R_i + R_j - \vec{r}_{ij}\hat{n}) -
               \gamma_n \vec{v}_{ij} \hat{n}
  \label{eq: fn}
\end{equation}
in the normal direction ($\hat{n}$) and
\begin{equation}
  F_{ij}^s = -\min(\gamma_s \vec v_{ij}\cdot\hat{s}(t), \mu|F_{ij}^n|) \ .
  \label{fric1}
\end{equation}
in the tangential direction ($\hat{s}$) of shearing. In Eq.~(\ref{eq: fn})
$\gamma_n$ represent the dynamic damping coefficient and Eq.~(\ref{fric1})
$\gamma_s$ represent the dynamic friction force in the tangential direction.
$\vec{r}_{ij}$ represents the vector  joining both centers of mass,
$\vec{v}_{ij}$  represents the relative motion of the two particles, and
$\tilde{Y}$ is related to the Young Modulus of the investigated material.
Dynamic friction for particle--particle collisions is defined in this model to
be proportional to the relative velocity of the particles in the tangential
direction which is a good approximation in many cases~\cite{schaefer96}.

During particle--wall contacts, the wall is treated as a particle with infinite
mass and radius. In the normal direction, Eq.~(\ref{eq: fn}) is applied,
whereas in the tangential direction, the static friction force
\begin{equation}
  \tilde{F}_{ij}^s = -\min(k_s \int \vec v_{ij}\cdot\hat{s}(t) dt,
  \mu|F_{ij}^n|)
  \label{fric2}
\end{equation}
is used. This is motivated by the observation that when particles flow along
the free surface, they dissipate most of their energy in collisions and can
come to rest in voids left by other particles. This is not possible at the 
cylinder walls.  In order to avoid additional artificial particles at the walls
we rather use a static friction law to avoid slipping and allowing for a static
surface angle when the rotation is stopped. Both tangential forces are limited
by the Coulomb criterion, see Eqs.~(\ref{fric1}) and (\ref{fric2}), which
states that the magnitude of the tangential force cannot exceed the magnitude
of the normal force multiplied by the friction coefficient $\mu$. The
coefficient of restitution for particle--particle collisions is set to 0.58
and  to 0.76 for particle--wall collisions. The large particles have a diameter
of 3~mm and a density of $\rho_l=1.3\frac{g}{cm^3}$. The material properties of
the large particles were chosen to correspond to the measured values of mustard
seeds~\cite{nakagawa93}. The small particles have a diameter of 2~mm. In order
to save computer time, we set $\tilde{Y}$ to \mbox{$8\cdot 10^3$ Pa~m} which is
about one order of magnitude softer than desired, but we checked that this has
no effect on the investigated properties of the material. This gives a contact
time during collisions of $8.5\cdot 10^{-5}$ s. The total number of particles
we used were up to 13 300.

\section{Dynamic Angle of Repose}
Using numerical simulations enables us to study arbitrary angle differences by
varying the inter-particle friction coefficient $\mu$ in Eqs.~(\ref{fric1})
and~(\ref{fric2}). For collisions between large particles, a value of
$\mu_l=0.2$ is used which gives a dynamic angle of repose similar to the
measured values of mustard seeds~\cite{dury97d}. When large particles touch the 
wall, a value of $\mu_w=0.4$ is used to avoid slipping at the boundary. The
additional friction at the end caps leads to an angle difference of 5$^\circ$
in our case which is in agreement with experiments~\cite{dury97d}. In order to
test different angle differences, the inter-particle friction coefficient for
the small particles, $\mu$, was varied  from 0.05 to 0.4. For collisions
between large and small particles, a value of $\mu_{\text{eff}} = \sqrt{\mu_l
\, \mu}$ is used.
\label{sec: angle}
\begin{figure}
  \resizebox{\hsize}{!}{%
    \includegraphics{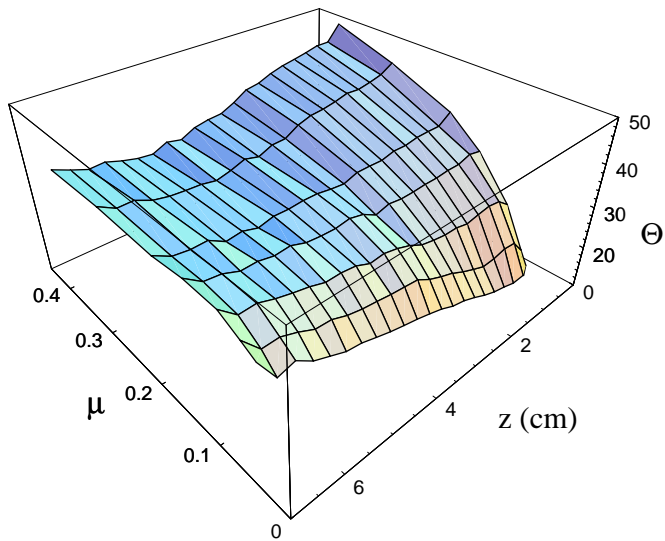}
  }
  \caption{Surface plot showing the dynamic angle of repose as function of 
           friction parameter $\mu$ and position along the rotational axis $z$.}
  \label{fig: conc_z}
\end{figure}

In Fig.~\ref{fig: conc_z}, we show the spatial variation of the dynamic angle
of repose, $\Theta$, and its dependence on $\mu$. The cylinder length was 7~cm
and the region initially occupied by small particles corresponds to the
interval $z=0\ldots 3.5$~cm to the right. The angles were measured by dividing
the cylinder into 22 slices along the rotational axis and we determined the
angle of repose for a rotation speed of 15~rpm via the center of mass of all
particles in each slice. In order to reduce the fluctuations, we averaged the
angles over an interval of 2 seconds after the first initial avalanches. When
$\mu$ is increased from 0.05 to 0.4, the measured angle at the right wall shows
a drastic increase from roughly 10$^\circ$ to 45$^\circ$. For glass beads it
was found that the dynamic angle of repose does hardly depend on the particle
size~\cite{zik94} and we achieve the same effect by using a value of $\mu=0.2$
in our numerical model. Also clearly visible are the effects of the two
cylinder end caps at $z=0$ and 7~cm, which lead to a higher angle due to the
additional wall friction and was studied in detail in Ref.~\cite{dury97d}.

\begin{figure}
  \resizebox{\hsize}{!}{%
    \includegraphics{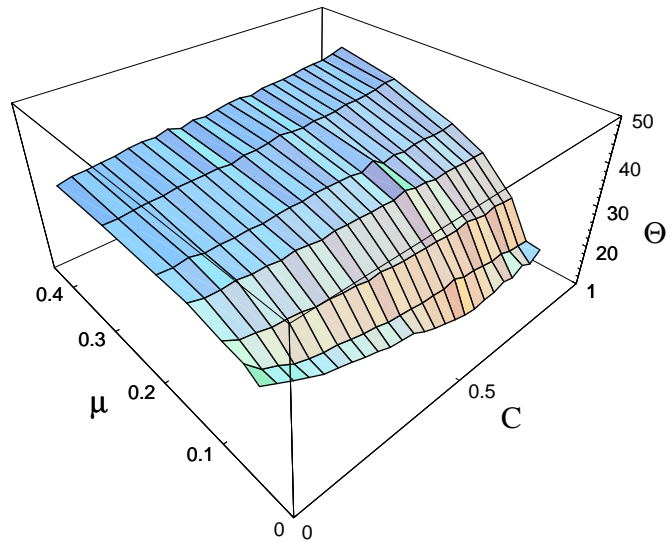}
  }
  \caption{Surface plot showing the dynamic angle of repose as function of 
           friction parameter $\mu$ and concentration of small particles $C$.}
  \label{fig: conc_c}
\end{figure}

From our numerical data, we can also calculate the concentration dependence of
the dynamic angle of repose. In order to reduce the influence of the boundary
caps, we only use the values for the angle of repose from the 16 central slices
of the total 22 slices and calculate the volumetric concentration of small
particles in each slice, denoted by $C$, which is shown in Fig.~\ref{fig:
conc_c} as function of the friction coefficient of the small particles,
$\mu$.   The graph shows the same general trend as Fig.~\ref{fig: conc_z} and
one can read off that no concentration dependence is observed for $\mu=0.2$
which agrees very well with an experimental study of 2 and 4~mm liquid-filled
spheres~\cite{hill97c}. In the same experiment, the concentration dependence
was investigated for a rotation speed of 30~rpm and it was found that the angle
increases with increasing concentration. Our numerical data clearly indicates
that the same concentration dependence can be found in our case when the small
particles have a higher friction coefficient than the large particles, see
e.g.\ the values for $\mu=0.4$ in Fig.~\ref{fig: conc_c}.

\section{Mixing at the Interface}
\label{sec: mixing}
\begin{figure}
  \resizebox{\hsize}{!}{%
    \includegraphics{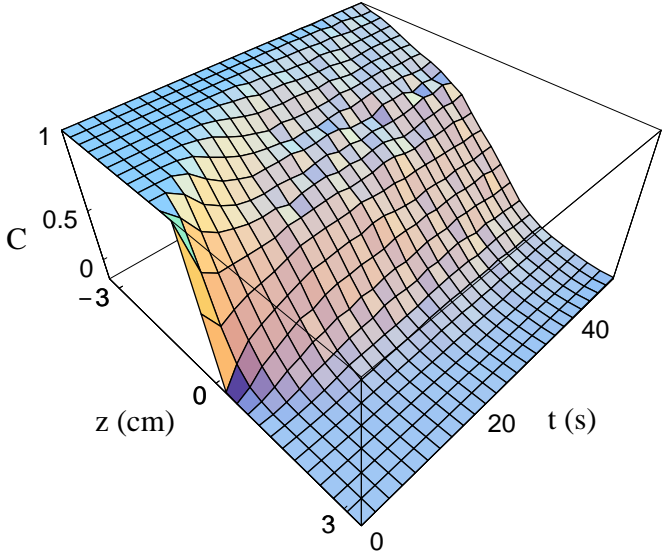}
  }
  \caption{Surface plot showing the time evolution of the concentration profile 
           for small particles along the rotational axis $z$.}
\label{fig: conc_t}
\end{figure}

The shape dynamics and the interface propagation of a binary particle mixture
was investigated for 1 and 4~mm liquid-filled spheres using magnetic resonance
imaging (MRI) \cite{ristow98}. The initial sharp interface between the regions
occupied by large and small particles, see Fig.~\ref{fig: sketch}, will deform
and move mostly due to particle diffusion in the fluidized surface layer. A
nearly fully segregated core of small particles was observed after rotating a
10~cm long, 7~cm wide cylinder for 10~min at 11.4~rpm. Since recording a full
three-dimensional MRI-image is still a time-consuming task and requires special
equipment, most studies divide the cylinder into vertical slices along the
rotational axis and record the particle concentration in each
slice~\cite{hogg66,cahn67,nakagawa97}. This leads to a one-dimensional
description of the mixing or segregation process and a typical example from our
numerical study is shown in Fig.~\ref{fig: conc_t}. The origin was shifted by
half the cylinder length to give a position of $z=0$~cm for the initial
interface which will be used throughout the rest of this chapter.   The
friction coefficient was $\mu=0.2$ and the density ratio $\rho/\rho_l=0.82$
where $\rho_l$ denotes the reference density of the large particles. The
initial sharp interface is clearly visible to the left and one notes how the
interface broadens in time. For $t=36$~s, the first small particles have
reached the right wall and consequently, the concentration values at the
boundaries will start to deviate from their initial values, already visible in
the profile to the far right for $t=50$~s.

\subsection{Approximation through pure Diffusion Process}
Assuming random particle motion in the axial direction ($z$ axis), one 
component systems could be well described by a diffusion process according to
Fick's Second Law~\cite{hogg66,cahn67}. The interface of a two component
system can also be studied in this fashion and the diffusion equation reads
\begin{equation}
  \frac{\partial C(z,t)}{\partial t} = \frac{\partial}{\partial z}
  \left( D \frac{\partial C(z,t)}{\partial z} \right)
  \label{Axial:eq_diff}
\end{equation}
where $C(z,t)$ and $D$ denote the relative concentration by volume of the 
smaller particles and the corresponding diffusion coefficient, respectively.
The initial condition for a cylinder with length $L$ are
\[
   C(z,0) = \left\{ \begin{array}{ll}
                     1, & \quad -\frac{L}{2} \le z < 0 \\
                     0, & \quad 0 < z \le \frac{L}{2}
                    \end{array} \right.
\]
whereas the boundary conditions read
\[ \left. \frac{\partial C}{\partial z} \right|_{z=-\frac{L}{2}} =
   \left. \frac{\partial C}{\partial z} \right|_{z=\frac{L}{2}} = 0 
\]
which states, that there is zero axial flux at the boundaries due to the end 
caps.

For a constant diffusion coefficient, Eq.~(\ref{Axial:eq_diff}) can be solved
analytically for the specified initial and boundary  conditions and the 
solution reads
\begin{eqnarray}
  C(z,t) = \frac{1}{2} - \frac{2}{\pi} \sum_{k=1}^\infty \frac{1}{2k-1}
  &\exp&\left(-\frac{(2k-1)^2\pi^2 D t}{L^2}\right)\times\nonumber\\ 
  &\sin&\left(\frac{(2k-1)\pi z}{L}\right) \ .
  \label{Axial:eq_profile}
\end{eqnarray}

In order to study the short time behaviour, we can solve our system by 
diffusion in an infinite cylinder. This is valid as long as the concentrations
at the real cylinder boundaries  have their initial values. Solving
Eq.~(\ref{Axial:eq_diff}) for this system  gives~\cite{crank75}
\begin{equation}
  C(z,t) = \frac{1}{2}\left[ 1-\text{Erf}\left(\frac{z}{2\sqrt{Dt}}\right) 
  \right]
\end{equation}
where $\text{Erf}(x)=\frac{2}{\sqrt{\pi}}\int_0^t e^{-t^2} dt$ is the error 
function. To determine now the diffusion coefficient we build the norm of
\begin{equation}
  {\mathfrak{A}}(t):=\left(C(.,t)-C(.,\infty)\right)\in \pmb{L}_2[0,\frac{L}{2}]
\end{equation}
where $C(z,\infty) = \frac{1}{2}$ is the steady state concentration:
\begin{align}
  ||\mathfrak{A}(t)||^2 &= \int_{-L/2}^0 \left(C(z,t)-C(z,\infty)\right)^2 dz 
  \notag \\
  &= \int_{-L/2}^0 \left( \text{Erf}\left(\frac{z}{2\sqrt{Dt}}\right)\right)^2 
  dz\label{eq: sigma} \\
  \intertext{which leads to}	
  \begin{split}&= \frac{L}{2}\text{Erf}^2\left(\frac{L}{4\sqrt{Dt}}\right)-
  \sqrt{\frac{Dt}{\pi}}
  \left\{
    -2\sqrt{2}\text{Erf}\left(\frac{L}{2\sqrt{2Dt}}\right) \right. \\
    &\quad\left.+4\exp\left(\frac{-L^2}{16Dt}\right)
    \text{Erf}\left(\frac{L}{4\sqrt{Dt}}\right)
  \right\}\ .\end{split}
\end{align}
But this result holds anyway just for small $t$ where \mbox{$C(L/2,t)\simeq 1$}
and therefore we have
\begin{align}
  \text{Erf}\left(\frac{L}{4\sqrt{Dt}}\right)&\simeq 1\notag \\
  \intertext{and out of the monotonic behavior of Erf() also}
  \text{Erf}\left(\frac{L}{2\sqrt{2Dt}}\right)&\simeq 1\notag \\
  \intertext{and for small $t$ we also have}
  \exp\left(-\frac{L^2}{16 D t}\right)&\simeq 0\notag \ .
\end{align}
Using this we finally get
\begin{equation}
  ||{\mathfrak{A}}(t)|| = ||{\mathfrak{A}}(0)|| \left( 
  1-\frac{4}{L}\sqrt{\frac{2Dt}{\pi}}
  \right) \text{ .}
  \label{eq: short_t}
\end{equation}

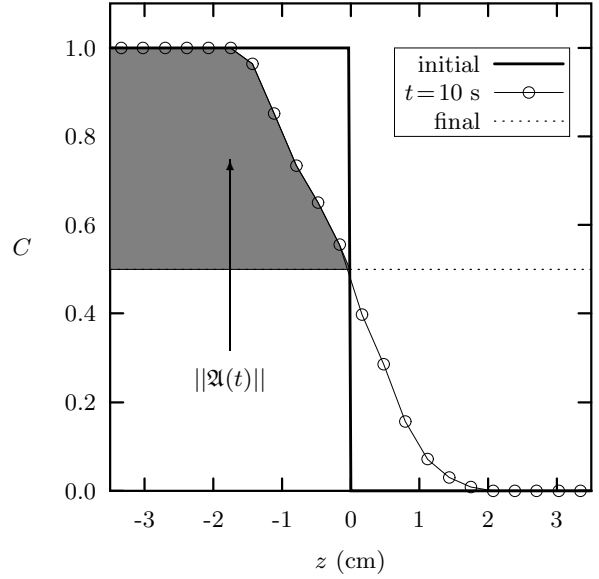
\begin{figure}
  {\centering
\setlength{\unitlength}{0.240900pt}
\begin{picture}(1500,920)(0,80)
\thicklines \path(199,179)(219,179)
\thicklines \path(954,179)(934,179)
\put(177,179){\makebox(0,0)[r]{0.0}}
\thicklines \path(199,318)(219,318)
\thicklines \path(954,318)(934,318)
\put(177,318){\makebox(0,0)[r]{0.2}}
\thicklines \path(199,458)(219,458)
\thicklines \path(954,458)(934,458)
\put(177,458){\makebox(0,0)[r]{0.4}}
\thicklines \path(199,597)(219,597)
\thicklines \path(954,597)(934,597)
\put(177,597){\makebox(0,0)[r]{0.6}}
\thicklines \path(199,736)(219,736)
\thicklines \path(954,736)(934,736)
\put(177,736){\makebox(0,0)[r]{0.8}}
\thicklines \path(199,875)(219,875)
\thicklines \path(954,875)(934,875)
\put(177,875){\makebox(0,0)[r]{1.0}}
\thicklines \path(253,179)(253,199)
\thicklines \path(253,945)(253,925)
\put(253,134){\makebox(0,0){-3}}
\thicklines \path(361,179)(361,199)
\thicklines \path(361,945)(361,925)
\put(361,134){\makebox(0,0){-2}}
\thicklines \path(469,179)(469,199)
\thicklines \path(469,945)(469,925)
\put(469,134){\makebox(0,0){-1}}
\thicklines \path(577,179)(577,199)
\thicklines \path(577,945)(577,925)
\put(577,134){\makebox(0,0){0}}
\thicklines \path(684,179)(684,199)
\thicklines \path(684,945)(684,925)
\put(684,134){\makebox(0,0){1}}
\thicklines \path(792,179)(792,199)
\thicklines \path(792,945)(792,925)
\put(792,134){\makebox(0,0){2}}
\thicklines \path(900,179)(900,199)
\thicklines \path(900,945)(900,925)
\put(900,134){\makebox(0,0){3}}
\thicklines \path(199,179)(954,179)(954,945)(199,945)(199,179)
\put(45,562){\makebox(0,0)[l]{\shortstack{$C$}}}
\put(576,67){\makebox(0,0){$z$ (cm)}}

\put(780,805){\makebox(0,0)[r]{$t\!=\!10$ s}}
\thinlines \path(802,805)(910,805)
\shade\path(199,875)(388,875)(422,850)(456,772)(491,690)(525,632)(559,566)(575,527)(199,527)(199,875)
\put(387,353){\makebox(0,0){$||{\mathfrak{A}}(t)||$}}
\put(387,400){\vector(0,1){300}}
\path(216,875)(250,875)(285,875)(319,875)(353,875)(388,875)(422,850)(456,772)(491,690)(525,632)(559,566)(594,456)(628,378)(662,288)(697,229)(731,200)(765,185)(800,179)(834,179)(868,179)(903,179)(937,179)
\put(216,875){\circle{18}}
\put(250,875){\circle{18}}
\put(285,875){\circle{18}}
\put(319,875){\circle{18}}
\put(353,875){\circle{18}}
\put(388,875){\circle{18}}
\put(422,850){\circle{18}}
\put(456,772){\circle{18}}
\put(491,690){\circle{18}}
\put(525,632){\circle{18}}
\put(559,566){\circle{18}}
\put(594,456){\circle{18}}
\put(628,378){\circle{18}}
\put(662,288){\circle{18}}
\put(697,229){\circle{18}}
\put(731,200){\circle{18}}
\put(765,185){\circle{18}}
\put(800,179){\circle{18}}
\put(834,179){\circle{18}}
\put(868,179){\circle{18}}
\put(903,179){\circle{18}}
\put(937,179){\circle{18}}
\put(856,805){\circle{18}}

\Thicklines
\put(780,850){\makebox(0,0)[r]{initial}}
\path(802,850)(910,850)
\path(199,875)(573,875)(577,179)(954,179)

\thinlines
\put(780,760){\makebox(0,0)[r]{final}}
\dottedline{13}(802,760)(910,760)
\dottedline{13}(199,527)(954,527)

\path(645,875)(920,875)(920,736)(645,736)(645,875)

\end{picture} }
  \caption{Concentration profiles for different times, from 
           Fig~\ref{fig: conc_t}.}
  \label{fig: profile}
\end{figure}

The physical interpretation of $||{\mathfrak{A}}(t)||$ will become clearer by
looking at a concentration profile extracted from  Fig.~\ref{fig: conc_t} which
is shown in Fig.~\ref{fig: profile}. Three profiles are shown, namely the
theoretical initial concentration profile as thick line, a computed profile for
$t=10$~s denoted by circles and the expected steady state profile as dotted
line. The quantity $||{\mathfrak{A}}(t)||$ is a measure of how close the
concentration profile is to the expected steady state profile and we shaded the
region which enters our calculations in Eq.~(\ref{eq: sigma}).

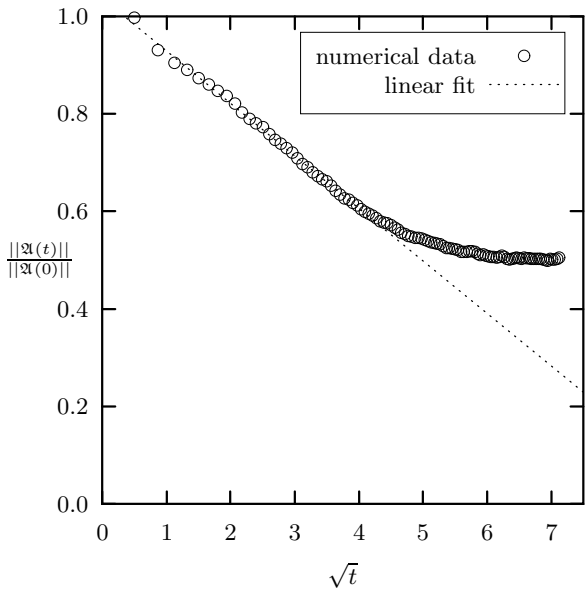
\begin{figure}
  {\centering
\setlength{\unitlength}{0.240900pt}
\begin{picture}(1500,920)(0,80)
\thicklines \path(199,179)(219,179)
\thicklines \path(954,179)(934,179)
\put(177,179){\makebox(0,0)[r]{0.0}}
\thicklines \path(199,332)(219,332)
\thicklines \path(954,332)(934,332)
\put(177,332){\makebox(0,0)[r]{0.2}}
\thicklines \path(199,485)(219,485)
\thicklines \path(954,485)(934,485)
\put(177,485){\makebox(0,0)[r]{0.4}}
\thicklines \path(199,639)(219,639)
\thicklines \path(954,639)(934,639)
\put(177,639){\makebox(0,0)[r]{0.6}}
\thicklines \path(199,792)(219,792)
\thicklines \path(954,792)(934,792)
\put(177,792){\makebox(0,0)[r]{0.8}}
\thicklines \path(199,945)(219,945)
\thicklines \path(954,945)(934,945)
\put(177,945){\makebox(0,0)[r]{1.0}}
\thicklines \path(199,179)(199,199)
\thicklines \path(199,945)(199,925)
\put(199,134){\makebox(0,0){0}}
\thicklines \path(300,179)(300,199)
\thicklines \path(300,945)(300,925)
\put(300,134){\makebox(0,0){1}}
\thicklines \path(400,179)(400,199)
\thicklines \path(400,945)(400,925)
\put(400,134){\makebox(0,0){2}}
\thicklines \path(501,179)(501,199)
\thicklines \path(501,945)(501,925)
\put(501,134){\makebox(0,0){3}}
\thicklines \path(602,179)(602,199)
\thicklines \path(602,945)(602,925)
\put(602,134){\makebox(0,0){4}}
\thicklines \path(702,179)(702,199)
\thicklines \path(702,945)(702,925)
\put(702,134){\makebox(0,0){5}}
\thicklines \path(803,179)(803,199)
\thicklines \path(803,945)(803,925)
\put(803,134){\makebox(0,0){6}}
\thicklines \path(904,179)(904,199)
\thicklines \path(904,945)(904,925)
\put(904,134){\makebox(0,0){7}}
\thicklines \path(199,179)(954,179)(954,945)(199,945)(199,179)
\put(45,562){\makebox(0,0)[l]{\shortstack{$\frac{||{\mathfrak{A}}(t)||}{||{\mathfrak{A}}(0)||}$}}}
\put(576,67){\makebox(0,0){$\sqrt{t}$}}

\thinlines
\put(249,943){\circle{18}}
\put(286,892){\circle{18}}
\put(312,872){\circle{18}}
\put(332,861){\circle{18}}
\put(350,848){\circle{18}}
\put(366,838){\circle{18}}
\put(380,828){\circle{18}}
\put(394,820){\circle{18}}
\put(407,808){\circle{18}}
\put(418,794){\circle{18}}
\put(430,784){\circle{18}}
\put(440,777){\circle{18}}
\put(451,771){\circle{18}}
\put(461,760){\circle{18}}
\put(470,751){\circle{18}}
\put(479,745){\circle{18}}
\put(488,738){\circle{18}}
\put(497,731){\circle{18}}
\put(505,722){\circle{18}}
\put(513,713){\circle{18}}
\put(521,708){\circle{18}}
\put(529,700){\circle{18}}
\put(537,694){\circle{18}}
\put(544,689){\circle{18}}
\put(551,686){\circle{18}}
\put(558,679){\circle{18}}
\put(565,671){\circle{18}}
\put(572,665){\circle{18}}
\put(579,659){\circle{18}}
\put(586,657){\circle{18}}
\put(592,652){\circle{18}}
\put(599,648){\circle{18}}
\put(605,642){\circle{18}}
\put(611,638){\circle{18}}
\put(617,635){\circle{18}}
\put(623,632){\circle{18}}
\put(629,628){\circle{18}}
\put(635,623){\circle{18}}
\put(641,621){\circle{18}}
\put(646,620){\circle{18}}
\put(652,617){\circle{18}}
\put(658,613){\circle{18}}
\put(663,610){\circle{18}}
\put(668,605){\circle{18}}
\put(674,603){\circle{18}}
\put(679,600){\circle{18}}
\put(684,599){\circle{18}}
\put(690,597){\circle{18}}
\put(695,597){\circle{18}}
\put(700,596){\circle{18}}
\put(705,594){\circle{18}}
\put(710,592){\circle{18}}
\put(715,590){\circle{18}}
\put(720,589){\circle{18}}
\put(724,588){\circle{18}}
\put(729,587){\circle{18}}
\put(734,584){\circle{18}}
\put(739,581){\circle{18}}
\put(743,581){\circle{18}}
\put(748,580){\circle{18}}
\put(753,579){\circle{18}}
\put(757,578){\circle{18}}
\put(762,575){\circle{18}}
\put(766,575){\circle{18}}
\put(771,575){\circle{18}}
\put(775,576){\circle{18}}
\put(779,576){\circle{18}}
\put(784,575){\circle{18}}
\put(788,572){\circle{18}}
\put(792,570){\circle{18}}
\put(797,571){\circle{18}}
\put(801,569){\circle{18}}
\put(805,568){\circle{18}}
\put(809,567){\circle{18}}
\put(813,567){\circle{18}}
\put(818,566){\circle{18}}
\put(822,567){\circle{18}}
\put(826,569){\circle{18}}
\put(830,567){\circle{18}}
\put(834,564){\circle{18}}
\put(838,563){\circle{18}}
\put(842,564){\circle{18}}
\put(846,565){\circle{18}}
\put(849,566){\circle{18}}
\put(853,565){\circle{18}}
\put(857,564){\circle{18}}
\put(861,566){\circle{18}}
\put(865,565){\circle{18}}
\put(869,565){\circle{18}}
\put(872,564){\circle{18}}
\put(876,564){\circle{18}}
\put(880,564){\circle{18}}
\put(884,564){\circle{18}}
\put(887,564){\circle{18}}
\put(891,563){\circle{18}}
\put(895,562){\circle{18}}
\put(898,561){\circle{18}}
\put(902,564){\circle{18}}
\put(905,563){\circle{18}}
\put(909,563){\circle{18}}
\put(913,564){\circle{18}}
\put(916,566){\circle{18}}

\dottedline{13}(237,942)(954,355)

\path(510,920)(925,920)(925,792)(510,792)(510,920)
\put(785,883){\makebox(0,0)[r]{numerical data}}
\put(861,883){\circle{18}}
\put(785,838){\makebox(0,0)[r]{linear fit}}
\dottedline{13}(807,838)(915,838)

\end{picture} }
  \caption{Plot of $\frac{||{\mathfrak{A}}(t)||}{||{\mathfrak{A}}(0)||}$ 
	   vs. $\sqrt{t}$. The linear fit, shown as dashed line, is used to
	   determine the diffusion coefficient.}
  \label{fig: sigma}
\end{figure}

The highest value of $||{\mathfrak{A}}(t)||$ is given for $t=0$~s and a
decrease linear in $\sqrt{t}$ is expected for short times, see Eq.~(\ref{eq:
short_t}). This is shown in Fig.~\ref{fig: sigma}, using the same simulation
parameters as for Figs.~\ref{fig: conc_t} and~\ref{fig: profile}, where we plot
$||{\mathfrak{A}}(t)||$ normalized by the initial value $||{\mathfrak{A}}(0)||$
vs.\ $\sqrt{t}$. From the slope of the linear fit shown as dotted line in
Fig.~\ref{fig: sigma}, we can calculate a constant diffusion coefficient based
on our approximations which gives $D=0.022 \pm 0.002$~cm$^2$/s and agrees well
with values extracted from experiments~\cite{ristow98}. When small particles
are close to the opposite wall, our approximation of an infinite long cylinder
does not hold anymore which leads to a systematic deviation from the
$\sqrt{t}$-behaviour, visible for times larger than 20~s in Fig.~\ref{fig:
sigma}.  For this specific run, the first small particle can be found in the
slice at the opoosite wall at $t=32$~s. This time difference of 15~s where the
graph deviates from the linear behaviour and the time the first particle
reaches the boundary comes from the fact that the particles feel the boundary
quite early.

\subsection{Dependence on Friction Coefficient}
\label{sec: friction}
\begin{figure}
  {\centering 
\setlength{\unitlength}{0.240900pt}
\begin{picture}(1500,920)(0,80)
\thicklines \path(243,179)(263,179)
\thicklines \path(954,179)(934,179)
\put(221,179){\makebox(0,0)[r]{0}}
\thicklines \path(243,256)(263,256)
\thicklines \path(954,256)(934,256)
\thicklines \path(243,332)(263,332)
\thicklines \path(954,332)(934,332)
\put(221,332){\makebox(0,0)[r]{0.01}}
\thicklines \path(243,409)(263,409)
\thicklines \path(954,409)(934,409)
\thicklines \path(243,485)(263,485)
\thicklines \path(954,485)(934,485)
\put(221,485){\makebox(0,0)[r]{0.02}}
\thicklines \path(243,562)(263,562)
\thicklines \path(954,562)(934,562)
\thicklines \path(243,639)(263,639)
\thicklines \path(954,639)(934,639)
\put(221,639){\makebox(0,0)[r]{0.03}}
\thicklines \path(243,715)(263,715)
\thicklines \path(954,715)(934,715)
\thicklines \path(243,792)(263,792)
\thicklines \path(954,792)(934,792)
\put(221,792){\makebox(0,0)[r]{0.04}}
\thicklines \path(243,868)(263,868)
\thicklines \path(954,868)(934,868)
\thicklines \path(243,945)(263,945)
\thicklines \path(954,945)(934,945)
\put(221,945){\makebox(0,0)[r]{0.05}}
\thicklines \path(243,179)(243,199)
\thicklines \path(243,945)(243,925)
\put(243,134){\makebox(0,0){0}}
\thicklines \path(322,179)(322,199)
\thicklines \path(322,945)(322,925)
\thicklines \path(401,179)(401,199)
\thicklines \path(401,945)(401,925)
\put(401,134){\makebox(0,0){0.1}}
\thicklines \path(480,179)(480,199)
\thicklines \path(480,945)(480,925)
\thicklines \path(559,179)(559,199)
\thicklines \path(559,945)(559,925)
\put(559,134){\makebox(0,0){0.2}}
\thicklines \path(638,179)(638,199)
\thicklines \path(638,945)(638,925)
\thicklines \path(717,179)(717,199)
\thicklines \path(717,945)(717,925)
\put(717,134){\makebox(0,0){0.3}}
\thicklines \path(796,179)(796,199)
\thicklines \path(796,945)(796,925)
\thicklines \path(875,179)(875,199)
\thicklines \path(875,945)(875,925)
\put(875,134){\makebox(0,0){0.4}}
\thicklines \path(954,179)(954,199)
\thicklines \path(954,945)(954,925)
\thicklines \path(243,179)(954,179)(954,945)(243,945)(243,179)
\put(45,562){\makebox(0,0)[l]{\shortstack{$D$ \\ 
$\left(\frac{\text{cm}^2}{\text{s}}\right)$}}}
\put(598,67){\makebox(0,0){$\mu$}}
\thinlines
\path(401,485)(401,853)
\path(391,485)(411,485)
\path(391,853)(411,853)
\path(480,531)(480,562)
\path(470,531)(490,531)
\path(470,562)(490,562)
\path(559,493)(559,508)
\path(549,493)(569,493)
\path(549,508)(569,508)
\path(717,393)(717,455)
\path(707,393)(727,393)
\path(707,455)(727,455)
\path(875,409)(875,424)
\path(865,409)(885,409)
\path(865,424)(885,424)
\put(401,669){\circle{18}}
\put(480,547){\circle{18}}
\put(559,501){\circle{18}}
\put(717,424){\circle{18}}
\put(875,416){\circle{18}}
\path(401,669)(401,669)(480,547)(559,501)(717,424)(875,416)
\end{picture} }
  \caption{Diffusion coefficient for different values of $\mu$ of the
	   small particles.}
\label{fig: dep_mu}
\end{figure}
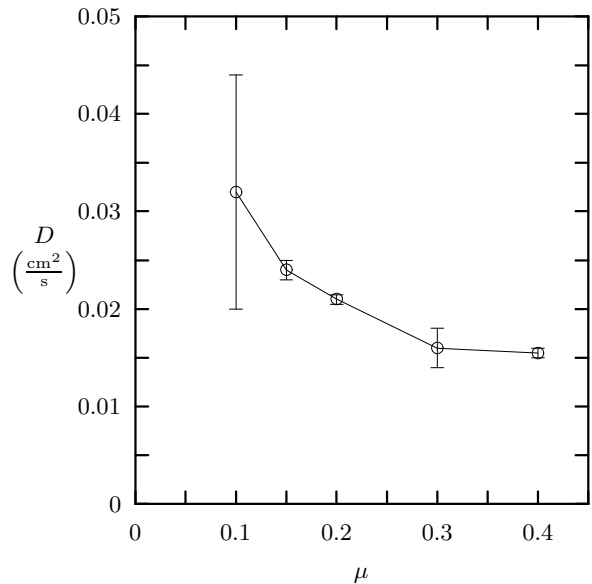

The dependence of the diffusion coefficient on the friction coefficient of the 
small particles is quite small and shown in Fig.~\ref{fig: dep_mu}. The
tendency of lower $D$ for higher $\mu$ even persists for quite large friction
coefficient, where small particles have a much higher angle of repose than the
large particles (for $\mu=0.2$  and $\Omega=15$~rpm the angle of repose is the
same for large and small  particles). This weak dependence can be explained by
the so called ``roller coaster'' effect. Suppose we have a sharp front between
small and large particles, then the angle of repose exhibits also a sharp
front. A particle on top of the free surface with the higher angle of repose
will see the angle difference and the motion of the particle will be directed
towards the region of the lower angle of repose. But the same thing happens in
the lower part of the free surface, where now the situation is reversed, and
the particle will move back. Therefore, in first approximation, there will be
no net effect on the drift (or diffusion) due to this difference in the angle
of repose, and what is left is a normal random walk on the free surface of the
particle in the direction along the rotational axis. If we now pay tribute to
the fact, that our particles have different sizes and therefore will exhibit
radial segregation, the ``roller coaster'' motion will not be as perfect as
described above. Suppose the small particles exhibit a higher angle of repose
($\mu>0.2$), the path of small particles will lead over the free surface of the
large particles and they can therefore be trapped into a radial core, thus will
be removed from the free surface motion and so also from the diffusion process,
which decreases the diffusion coefficient $D$. On the other hand for $\mu<0.2$
, large particle will not be trapped into a core and can continue to
participate in the diffusional process even when they get stuck during the
``roller coaster''  motion, which is more probable the wider the ``roller
coaster'' path, i.e. the larger the  difference in the angle of repose. So $D$
increases with decreasing $\mu$.

\begin{figure}
  \begin{center}
    \hbox{\resizebox{0.49\hsize}{!}{%
            \includegraphics{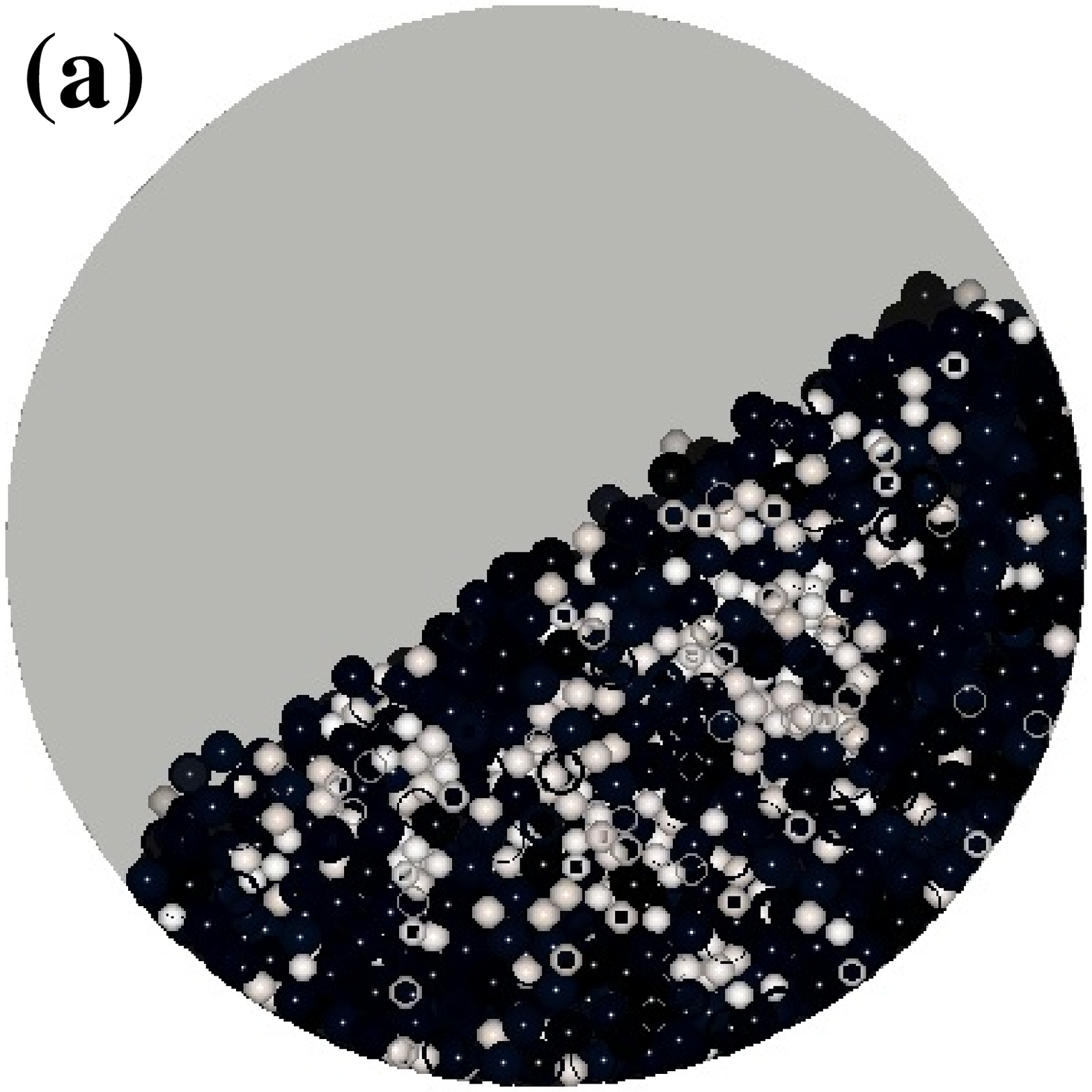}
	  } \hfill
	  \resizebox{0.49\hsize}{!}{%
            \includegraphics{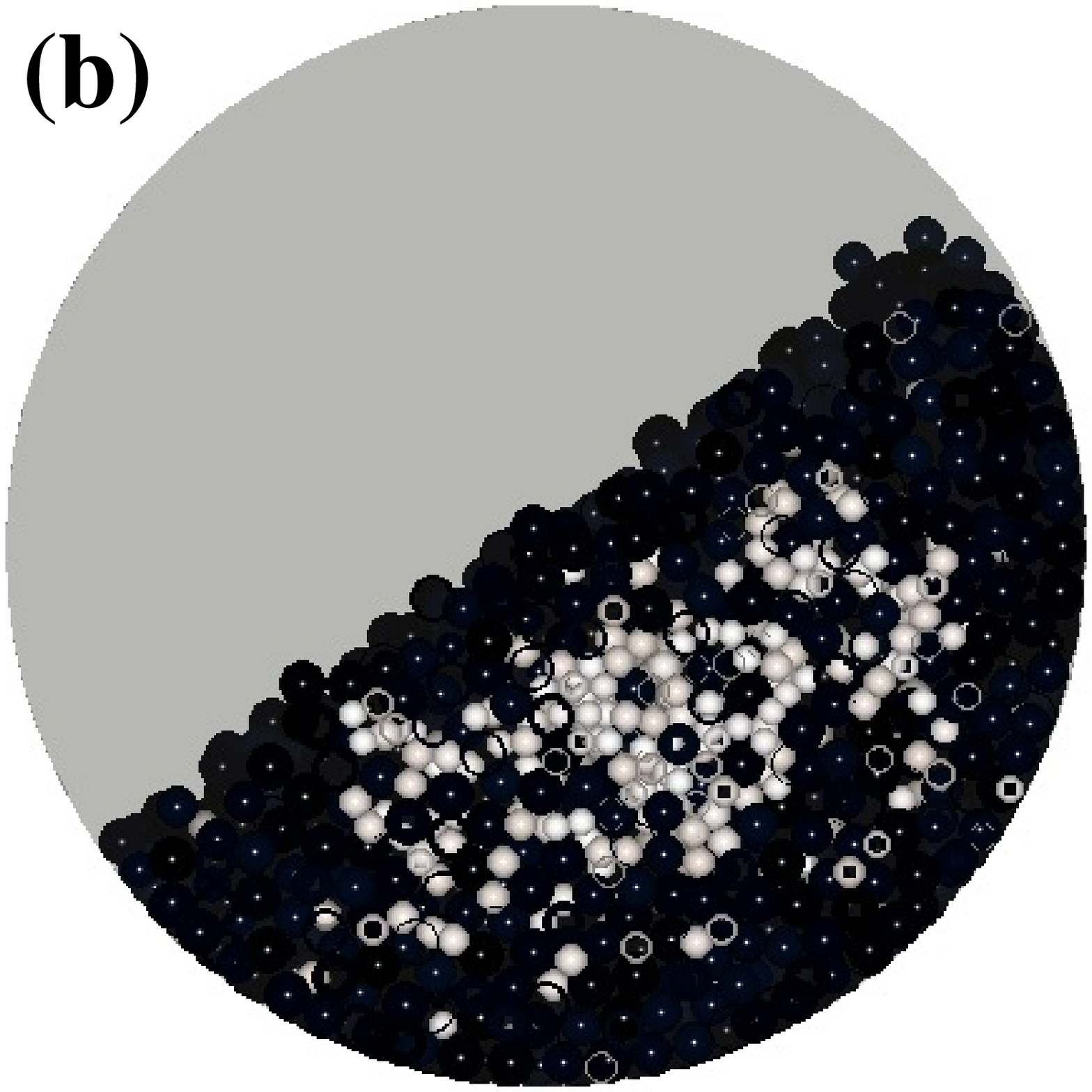}
	  }
    }
  \end{center}
  \caption{Cross section of the cylinder close to the inital interface for (a)
           $\mu=0.15$ at $t=33$ s and (b) $\mu=0.4$ at $t=27$~s. Large
           particles are shown in black and small particles in white.}
  \label{fig: axial_08}
\end{figure}

In order to demonstrate the increase of radial segregation due to an increase
of friction coefficient, we show in Fig.~\ref{fig: axial_08} two cross sections
of the cylinder which represent the configuration close to the initial
interface. For a value of $\mu=0.15$, we show in part (a) the particle
configuration at $t=33$~s and a segregation of the small particles, drawn in
white, is hardly visible. In contrast to this, a nice segregation is visible in
part (b), which shows a configuration for $\mu=0.4$ and $t=27$~s. This supports
our hypothesis that radial segregation will {\em hinder} the diffusion of small
particles and thus decrease the diffusion coefficient with increasing friction
coefficient.

\subsection{Dependence on Rotation Speed}
\label{sec: omega}
When investigating the mixing process of glass beads, Hogg et al.~\cite{hogg66}
found that the dynamics could be well described by using the number of
revolutions instead of the time in Eq.~(\ref{Axial:eq_diff}). This directly
implies that the calculated diffusion constant should be directly proportional
to the rotation speed of the cylinder. We checked this for our system by
investigating an $\Omega$-range of 7.5 to 45~rpm for the simulation parameters
$\rho/\rho_l=1$ and $\mu=0.15$ which is shown in Fig.~\ref{Axial:diff_omega}.
Also shown as dashed line is the linear dependence proposed in~\cite{hogg66}
and as expected it is only a valid assumption for low rotation speeds. On the
other hand, our numerically calculated values for $D$ rather show a more than
linear dependence when the whole $\Omega$-range is considered, which was fitted
by a quadratic function and added as a solid line to
Fig.~\ref{Axial:diff_omega}.  This deviation from the linear behavior is due to
the fact that the particles will bounce off the cylinder wall, after they flowed
down the free surface. This effect of bouncing is also observed in
experiments~\cite{nityanand86}.

\subsection{Dependence on Density Ratio}
\label{sec: density}
\begin{figure*}
   \hbox{ 
\setlength{\unitlength}{0.240900pt}
\begin{picture}(1500,920)(0,80)
\thicklines \path(199,179)(219,179)
\thicklines \path(954,179)(934,179)
\put(177,179){\makebox(0,0)[r]{0.0}}
\thicklines \path(199,275)(219,275)
\thicklines \path(954,275)(934,275)
\thicklines \path(199,371)(219,371)
\thicklines \path(954,371)(934,371)
\put(177,371){\makebox(0,0)[r]{0.4}}
\thicklines \path(199,466)(219,466)
\thicklines \path(954,466)(934,466)
\thicklines \path(199,562)(219,562)
\thicklines \path(954,562)(934,562)
\put(177,562){\makebox(0,0)[r]{0.8}}
\thicklines \path(199,658)(219,658)
\thicklines \path(954,658)(934,658)
\thicklines \path(199,754)(219,754)
\thicklines \path(954,754)(934,754)
\put(177,754){\makebox(0,0)[r]{1.2}}
\thicklines \path(199,849)(219,849)
\thicklines \path(954,849)(934,849)
\thicklines \path(199,945)(219,945)
\thicklines \path(954,945)(934,945)
\put(177,945){\makebox(0,0)[r]{1.6}}
\thicklines \path(199,179)(199,199)
\thicklines \path(199,945)(199,925)
\put(199,134){\makebox(0,0){0}}
\thicklines \path(291,179)(291,199)
\thicklines \path(291,945)(291,925)
\thicklines \path(383,179)(383,199)
\thicklines \path(383,945)(383,925)
\put(383,134){\makebox(0,0){1}}
\thicklines \path(475,179)(475,199)
\thicklines \path(475,945)(475,925)
\thicklines \path(567,179)(567,199)
\thicklines \path(567,945)(567,925)
\put(567,134){\makebox(0,0){2}}
\thicklines \path(659,179)(659,199)
\thicklines \path(659,945)(659,925)
\thicklines \path(751,179)(751,199)
\thicklines \path(751,945)(751,925)
\put(751,134){\makebox(0,0){3}}
\thicklines \path(844,179)(844,199)
\thicklines \path(844,945)(844,925)
\thicklines \path(936,179)(936,199)
\thicklines \path(936,945)(936,925)
\put(936,134){\makebox(0,0){4}}
\thicklines \path(199,179)(954,179)(954,945)(199,945)(199,179)
\put(45,658){\makebox(0,0)[l]{\shortstack{$D$ \\
$\left(\frac{\text{cm}^2}{\text{s}}\right)$}}}
\put(576,67){\makebox(0,0){$\rho/\rho_l$}}
\put(844,849){\makebox(0,0){\large\bf (a)}}
\thinlines
\path(217,739)(217,864)
\path(207,739)(227,739)
\path(207,864)(227,864)
\path(245,427)(245,482)
\path(235,427)(255,427)
\path(235,482)(255,482)
\path(291,254)(291,270)
\path(281,254)(301,254)
\path(281,270)(301,270)
\path(337,201)(337,206)
\path(327,201)(347,201)
\path(327,206)(347,206)
\path(350,188)(350,191)
\path(340,188)(360,188)
\path(340,191)(360,191)
\path(383,183)(383,184)
\path(373,183)(393,183)
\path(373,184)(393,184)
\path(429,203)(429,208)
\path(419,203)(439,203)
\path(419,208)(439,208)
\path(475,225)(475,235)
\path(465,225)(485,225)
\path(465,235)(485,235)
\path(567,262)(567,281)
\path(557,262)(577,262)
\path(557,281)(577,281)
\path(751,317)(751,348)
\path(741,317)(761,317)
\path(741,348)(761,348)
\path(936,382)(936,427)
\path(926,382)(946,382)
\path(926,427)(946,427)
\put(217,801){\circle{18}}
\put(245,455){\circle{18}}
\put(291,262){\circle{18}}
\put(337,204){\circle{18}}
\put(350,190){\circle{18}}
\put(383,184){\circle{18}}
\put(429,206){\circle{18}}
\put(475,230){\circle{18}}
\put(567,271){\circle{18}}
\put(751,332){\circle{18}}
\put(936,404){\circle{18}}
\path(217,801)(217,801)(245,455)(291,262)(337,204)(350,190)(383,184)(429,206)(475,230)(567,271)(751,332)(936,404)
\end{picture} \hspace{-9.7cm} 
\setlength{\unitlength}{0.175pt}
\begin{picture}(1500,900)(0,-465)
\thicklines \path(154,112)(174,112)
\thicklines \path(742,112)(722,112)
\put(132,112){\makebox(0,0)[r]{\small 0.0}}
\thicklines \path(154,196)(174,196)
\thicklines \path(742,196)(722,196)
\put(132,196){\makebox(0,0)[r]{}}
\thicklines \path(154,279)(174,279)
\thicklines \path(742,279)(722,279)
\put(132,279){\makebox(0,0)[r]{\small 0.1}}
\thicklines \path(154,363)(174,363)
\thicklines \path(742,363)(722,363)
\put(132,363){\makebox(0,0)[r]{}}
\thicklines \path(154,446)(174,446)
\thicklines \path(742,446)(722,446)
\put(132,446){\makebox(0,0)[r]{\small 0.2}}
\thicklines \path(154,530)(174,530)
\thicklines \path(742,530)(722,530)
\put(132,530){\makebox(0,0)[r]{}}
\thicklines \path(154,613)(174,613)
\thicklines \path(742,613)(722,613)
\put(132,613){\makebox(0,0)[r]{\small 0.3}}
\thicklines \path(203,112)(203,132)
\thicklines \path(203,697)(203,677)
\put(203,67){\makebox(0,0){\small 0.5}}
\thicklines \path(326,112)(326,132)
\thicklines \path(326,697)(326,677)
\put(326,67){\makebox(0,0){\small 0.75}}
\thicklines \path(448,112)(448,132)
\thicklines \path(448,697)(448,677)
\put(448,67){\makebox(0,0){\small 1.0}}
\thicklines \path(571,112)(571,132)
\thicklines \path(571,697)(571,677)
\put(571,67){\makebox(0,0){\small 1.25}}
\thicklines \path(693,112)(693,132)
\thicklines \path(693,697)(693,677)
\put(693,67){\makebox(0,0){\small 1.5}}
\thicklines \path(154,112)(742,112)(742,697)(154,697)(154,112)
\thinlines \path(203,371)(203,431)
\thinlines \path(193,371)(213,371)
\thinlines \path(193,431)(213,431)
\thinlines \path(326,191)(326,207)
\thinlines \path(316,191)(336,191)
\thinlines \path(316,207)(336,207)
\thinlines \path(360,144)(360,154)
\thinlines \path(350,144)(370,144)
\thinlines \path(350,154)(370,154)
\thinlines \path(448,127)(448,130)
\thinlines \path(438,127)(458,127)
\thinlines \path(438,130)(458,130)
\thinlines \path(571,197)(571,214)
\thinlines \path(561,197)(581,197)
\thinlines \path(561,214)(581,214)
\thinlines \path(693,274)(693,308)
\thinlines \path(683,274)(703,274)
\thinlines \path(683,308)(703,308)
\put(203,401){\circle{18}}
\put(326,199){\circle{18}}
\put(360,149){\circle{18}}
\put(448,129){\circle{18}}
\put(571,206){\circle{18}}
\put(693,291){\circle{18}}
\thinlines \path(154,672)(154,672)(160,641)(166,611)(172,583)(178,556)(184,530)(190,505)(196,481)(202,458)(207,437)(213,416)(219,396)(225,378)(231,360)(237,343)(243,327)(249,312)(255,297)(261,284)(267,271)(273,258)(279,247)(285,236)(291,226)(297,216)(302,207)(308,199)(314,191)(320,184)(326,177)(332,171)(338,165)(344,160)(350,155)(356,151)(362,147)(368,144)(374,141)(380,138)(386,136)(392,134)(398,132)(403,131)(409,130)(415,129)(421,129)(427,128)(433,129)(439,129)(445,130)
\thinlines \path(445,130)(451,131)(457,132)(463,133)(469,135)(475,136)(481,138)(487,141)(493,143)(498,145)(504,148)(510,151)(516,154)(522,157)(528,160)(534,163)(540,167)(546,170)(552,174)(558,178)(564,182)(570,186)(576,190)(582,194)(588,198)(594,202)(599,207)(605,211)(611,216)(617,220)(623,225)(629,230)(635,234)(641,239)(647,244)(653,248)(659,253)(665,258)(671,263)(677,268)(683,273)(689,278)(694,283)(700,288)(706,293)(712,298)(718,303)(724,308)(730,313)(736,317)(742,322)
\end{picture}
          \hspace{-3.4cm} 
\setlength{\unitlength}{0.240900pt}
\begin{picture}(1500,920)(0,80)
\thicklines \path(199,179)(219,179)
\thicklines \path(954,179)(934,179)
\put(177,179){\makebox(0,0)[r]{0}}
\thicklines \path(199,288)(219,288)
\thicklines \path(954,288)(934,288)
\put(177,288){\makebox(0,0)[r]{0.1}}
\thicklines \path(199,398)(219,398)
\thicklines \path(954,398)(934,398)
\put(177,398){\makebox(0,0)[r]{0.2}}
\thicklines \path(199,507)(219,507)
\thicklines \path(954,507)(934,507)
\put(177,507){\makebox(0,0)[r]{0.3}}
\thicklines \path(199,617)(219,617)
\thicklines \path(954,617)(934,617)
\put(177,617){\makebox(0,0)[r]{0.4}}
\thicklines \path(199,726)(219,726)
\thicklines \path(954,726)(934,726)
\put(177,726){\makebox(0,0)[r]{0.5}}
\thicklines \path(199,836)(219,836)
\thicklines \path(954,836)(934,836)
\put(177,836){\makebox(0,0)[r]{0.6}}
\thicklines \path(199,945)(219,945)
\thicklines \path(954,945)(934,945)
\put(177,945){\makebox(0,0)[r]{0.7}}
\thicklines \path(199,179)(199,199)
\thicklines \path(199,945)(199,925)
\put(199,134){\makebox(0,0){0}}
\thicklines \path(283,179)(283,199)
\thicklines \path(283,945)(283,925)
\thicklines \path(367,179)(367,199)
\thicklines \path(367,945)(367,925)
\put(367,134){\makebox(0,0){1}}
\thicklines \path(451,179)(451,199)
\thicklines \path(451,945)(451,925)
\thicklines \path(535,179)(535,199)
\thicklines \path(535,945)(535,925)
\put(535,134){\makebox(0,0){2}}
\thicklines \path(618,179)(618,199)
\thicklines \path(618,945)(618,925)
\thicklines \path(702,179)(702,199)
\thicklines \path(702,945)(702,925)
\put(702,134){\makebox(0,0){3}}
\thicklines \path(786,179)(786,199)
\thicklines \path(786,945)(786,925)
\thicklines \path(870,179)(870,199)
\thicklines \path(870,945)(870,925)
\put(870,134){\makebox(0,0){4}}
\thicklines \path(954,179)(954,199)
\thicklines \path(954,945)(954,925)
\thicklines \path(199,179)(954,179)(954,945)(199,945)(199,179)
\put(45,562){\makebox(0,0)[l]{\shortstack{$q_\infty$}}}
\put(576,67){\makebox(0,0){$\rho/\rho_l$}}
\put(844,288){\makebox(0,0){\large\bf (b)}}
\thinlines
\path(870,792)(870,814)
\path(860,792)(880,792)
\path(860,814)(880,814)
\path(702,814)(702,836)
\path(692,814)(712,814)
\path(692,836)(712,836)
\path(535,825)(535,868)
\path(525,825)(545,825)
\path(525,868)(545,868)
\path(451,836)(451,857)
\path(441,836)(461,836)
\path(441,857)(461,857)
\path(409,814)(409,857)
\path(399,814)(419,814)
\path(399,857)(419,857)
\path(367,726)(367,770)
\path(357,726)(377,726)
\path(357,770)(377,770)
\path(337,343)(337,781)
\path(327,343)(347,343)
\path(327,781)(347,781)
\path(325,234)(325,343)
\path(315,234)(335,234)
\path(315,343)(335,343)
\path(283,190)(283,212)
\path(273,190)(293,190)
\path(273,212)(293,212)
\path(241,179)(241,223)
\path(231,179)(251,179)
\path(231,223)(251,223)
\put(870,803){\circle{18}}
\put(702,825){\circle{18}}
\put(535,847){\circle{18}}
\put(451,847){\circle{18}}
\put(409,836){\circle{18}}
\put(367,748){\circle{18}}
\put(337,562){\circle{18}}
\put(325,288){\circle{18}}
\put(283,201){\circle{18}}
\put(241,201){\circle{18}}
\path(870,803)(702,825)(535,847)(451,847)(409,836)(367,748)(337,562)(325,288)(283,201)(241,201)
\end{picture} }
   \vspace{2ex}
   \caption{Calculation of the functional dependence on the density ratio 
            $\frac{\rho}{\rho_l}$ of the small particles for (a) diffusion 
	    coefficient, $D$, where the inset magnifies the region close to
	    $\frac{\rho}{\rho_l}=1$ with a non-linear fit as solid line and (b) 
	    final amount of segregation, $q_\infty$.}
   \label{fig: dep_rho}
\end{figure*}
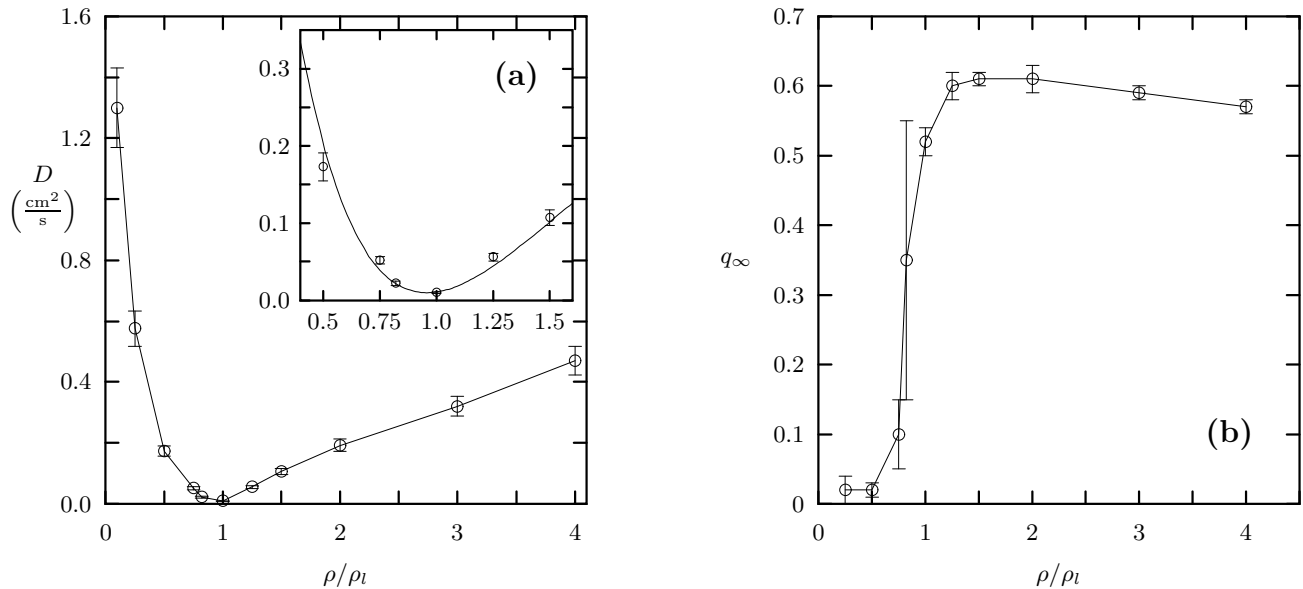

The particle motion depends on the density ratio $\rho/\rho_l$ which is
illustrated in Fig.~\ref{fig: dep_rho} for a constant value of $\mu=0.2$. To 
the left, Fig.~\ref{fig: dep_rho}a, the diffusion constant is plotted as 
function of this density ratio showing a minimum value for $\rho/\rho_l=1$ and
a large increase for lower and higher valus. In contrast to the previous
section, radial segregation will be present towards both sides of the graph. In
general, smaller and denser particles will segregate radially, so increasing
the density ratio will enhance radial segregation, but when decreasing the
density ratio, the larger particles become denser and eventually the large
particles will segregate into the radial core. Also shown in the same graph as
inset is a magnification of the region close to $\frac{\rho}{\rho_l}=1$ with a 
non-linear fit as solid line. This inset shows that our numerical model always
gives a diffusion coefficient larger than zero thus indicating that the front is
{\em not} stable, regardless of the density ratio of the two particle
components. 

In order to quantify this process, we use a procedure outlined in
Refs.~\cite{dury97,dury98} to normalize the final amount of segregation. The
drum is divided into concentric rings and the final percentage by volume of 
small particles for large times is estimated in each ring and normalized with
respect to a perfectly well radially segregated configuration. Due to the
non-negligible width of the fluidized layer, a value of one cannot be achieved.
This quantity, denoted by $q_\infty$, is plotted in Fig.~\ref{fig: dep_rho}b as
function of the density ratio. With increasing density ratio, $q_\infty$
increases and saturates around a value of 0.6. The slight decrease for values
of $\rho/\rho_l>2$ is due to the definition of $q_\infty$ which does not take
the shifting of the center of mass of the smaller particles into account. For
$\rho/\rho_l=0.5$, the two competing effects of size- and mass-segregation
cancel each other and we get a perfect mixing of small and large particles
indicated by a small value of $q_\infty$ in Fig.~\ref{fig: dep_rho}b. 

\begin{figure}	
  {\centering 
\setlength{\unitlength}{0.240900pt}
\begin{picture}(1500,920)(0,80)
\thicklines \path(221,179)(241,179)
\thicklines \path(954,179)(934,179)
\put(199,179){\makebox(0,0)[r]{0}}
\thicklines \path(221,307)(241,307)
\thicklines \path(954,307)(934,307)
\thicklines \path(221,434)(241,434)
\thicklines \path(954,434)(934,434)
\put(199,434){\makebox(0,0)[r]{0.02}}
\thicklines \path(221,562)(241,562)
\thicklines \path(954,562)(934,562)
\thicklines \path(221,690)(241,690)
\thicklines \path(954,690)(934,690)
\put(199,690){\makebox(0,0)[r]{0.04}}
\thicklines \path(221,817)(241,817)
\thicklines \path(954,817)(934,817)
\thicklines \path(221,945)(241,945)
\thicklines \path(954,945)(934,945)
\put(199,945){\makebox(0,0)[r]{0.06}}
\thicklines \path(221,179)(221,199)
\thicklines \path(221,945)(221,925)
\put(221,134){\makebox(0,0){0}}
\thicklines \path(294,179)(294,199)
\thicklines \path(294,945)(294,925)
\thicklines \path(368,179)(368,199)
\thicklines \path(368,945)(368,925)
\put(368,134){\makebox(0,0){10}}
\thicklines \path(441,179)(441,199)
\thicklines \path(441,945)(441,925)
\thicklines \path(514,179)(514,199)
\thicklines \path(514,945)(514,925)
\put(514,134){\makebox(0,0){20}}
\thicklines \path(588,179)(588,199)
\thicklines \path(588,945)(588,925)
\thicklines \path(661,179)(661,199)
\thicklines \path(661,945)(661,925)
\put(661,134){\makebox(0,0){30}}
\thicklines \path(734,179)(734,199)
\thicklines \path(734,945)(734,925)
\thicklines \path(807,179)(807,199)
\thicklines \path(807,945)(807,925)
\put(807,134){\makebox(0,0){40}}
\thicklines \path(881,179)(881,199)
\thicklines \path(881,945)(881,925)
\thicklines \path(954,179)(954,199)
\thicklines \path(954,945)(954,925)
\put(954,134){\makebox(0,0){50}}
\thicklines \path(221,179)(954,179)(954,945)(221,945)(221,179)
\put(45,562){\makebox(0,0)[l]{\shortstack{$D$ \\ 
$\left(\frac{\text{cm}^2}{\text{s}}\right)$}}}
\put(587,67){\makebox(0,0){$\Omega$ (rpm)}}

\thinlines
\dottedline{13}(221,179)(954,710)
\path(221,179)(221,180)
\path(211,179)(231,179)
\path(211,180)(231,180)
\path(332,256)(332,266)
\path(322,256)(342,256)
\path(322,266)(342,266)
\path(441,332)(441,342)
\path(431,332)(451,332)
\path(431,342)(451,342)
\path(551,466)(551,492)
\path(541,466)(561,466)
\path(541,492)(561,492)
\path(661,562)(661,639)
\path(651,562)(671,562)
\path(651,639)(671,639)
\path(771,702)(771,728)
\path(761,702)(781,702)
\path(761,728)(781,728)
\path(880,862)(880,913)
\path(870,862)(890,862)
\path(870,913)(890,913)
\put(221,179){\circle{18}}
\put(332,261){\circle{18}}
\put(441,337){\circle{18}}
\put(551,479){\circle{18}}
\put(661,600){\circle{18}}
\put(771,715){\circle{18}}
\put(880,888){\circle{18}}
\path(221,179)(221,179)(228,184)(236,188)(243,193)(251,198)(258,202)(265,207)(273,212)(280,217)(288,222)(295,228)(302,233)(310,238)(317,244)(325,249)(332,255)(339,260)(347,266)(354,272)(362,278)(369,284)(376,290)(384,296)(391,302)(399,308)(406,315)(414,321)(421,328)(428,334)(436,341)(443,348)(451,354)(458,361)(465,368)(473,375)(480,382)(488,389)(495,397)(502,404)(510,411)(517,419)(525,426)(532,434)(539,442)(547,450)(554,457)(562,465)(569,473)(576,482)(584,490)
\path(584,490)(591,498)(599,506)(606,515)(613,523)(621,532)(628,540)(636,549)(643,558)(650,567)(658,576)(665,585)(673,594)(680,603)(687,612)(695,621)(702,631)(710,640)(717,650)(724,659)(732,669)(739,679)(747,689)(754,699)(761,709)(769,719)(776,729)(784,739)(791,749)(799,760)(806,770)(813,781)(821,791)(828,802)(836,813)(843,824)(850,835)(858,846)(865,857)(873,868)(880,879)(887,890)(895,902)(902,913)(910,925)(917,936)(922,945)

\path(540,370)(930,370)(930,240)(540,240)(540,370)
\put(807,347){\makebox(0,0)[r]{numerical data}}
\put(870,347){\circle{18}}
\put(807,307){\makebox(0,0)[r]{linear fit}}
\dottedline{13}(830,307)(910,307)
\put(807,267){\makebox(0,0)[r]{quadratic fit}}
\path(830,267)(910,267)

\end{picture} }
  \caption{Diffusion coefficient for different values of the angular velocity 
           $\Omega$ of the cylinder.}
  \label{Axial:diff_omega}
\end{figure}
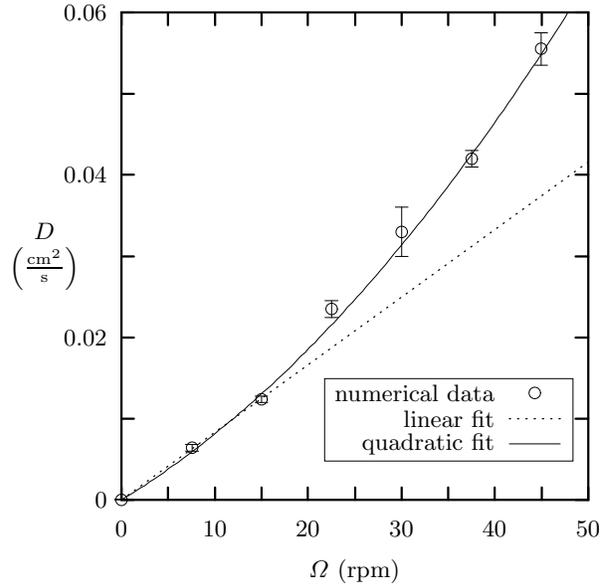

Remembering our hypothesis that radial segregation will hinder diffusion from
Sec.~\ref{sec: friction}, one might wonder why the diffusion coefficient starts
to rise again for $\rho/\rho_l>1$ where we get better segregation, whereas for 
$\rho/\rho_l<1$ the diffusion coefficient rises as expected. In contrast to the
previous section, we now have to take different densities for the particles into
account. We are starting with an initial front where the small particles are on
the left and  the large particles on the right half of the cylinder (see
Fig.~\ref{fig: sketch}).  The  pressure at the interface resulting from the
particles above is in first approximation the hydrostatic pressure for granular
media~\cite{wieghardt74}:
\begin{equation}
  p =  c \rho g \left(1-e^{\frac{-h_d}{c}}\right)
  \label{eq: pressure}
\end{equation}
where c is a parameter, which depends on the friction coefficient and the 
boundaries of the investigated geometry. The initial pressure for small depths
$h_d$ is like in a fluid $p=\rho g h_d$, but for larger $h_d$ the pressure
saturates  exponentially to $p=\rho g c$, in contrast to all normal liquids.
This general pressure dependence is by itself an interesting property for
granular media and is independent of the grain size.

For density ratios different than $1$, we get a pressure difference at the 
interface which enhances the mixing of the particles. This results in our
simple model in a higher diffusion coefficient and consequently, we get a
minimum in $D$ for $\rho/\rho_l=1$ which is clearly visible in Fig.~\ref{fig:
dep_rho}a. This phenomena will be discussed in more detail in section~\ref{sec:
drift} below.

\section{Microscopic Calculation of Diffusion Coefficient}
\label{sec: macro}
Another way to investigate the front diffusion is by looking at the
particle trajectories directly. When recording only the displacement of the
particles along the rotational axis, we obtain a {\em microscopic} definition
of a diffusion coefficient, $D_m$, through
\begin{equation}
  \langle z(t)^2 \rangle -\langle z(t) \rangle^2 = 2 D_m t +\langle z(0) 
  \rangle^2 
  \text{ ,}
\end{equation}
where the spatial average is done over all $N$ investigated particles via
\begin{equation}
  \langle z(t) \rangle := \frac{1}{N}\sum_{i=1}^{N} \ z_i(t) 
  \text{ .}
\end{equation}
This technique to obtain a microscopic diffusion constant is similar to the one 
used in the experiments by Zik and Stavans~\cite{zik91} who investigated 
the diffusional behaviour of vertically shaken granular material. It enables us
to explicitly consider a drift of the particles whereas this was incorporated
into the macroscopic diffusion constant presented in the previous
section. The drift velocity, $v$, is defined by the following relation:
\begin{equation}
  \langle z(t) \rangle -\langle z(0) \rangle = vt\text{ .} 
  \label{eq: drift}
\end{equation}
Another possibility to obtain a microscopic diffusion coefficient would be
\[
  \langle \left(z(t)-z(0)\right)^2 \rangle = 2D_m't \text{ ,}
\]
but in this case, one would also absorb a possible drift into the diffusion 
coefficient. It is therefore only useful for pure diffusion processes and thus
will not be used further in this section, because we want to investigate the
drift velocity and the diffusion coefficient separately.

\subsection{Diffusion Coefficient}
\begin{figure*}
  \hbox{ 
\setlength{\unitlength}{0.240900pt}
\begin{picture}(1500,920)(0,80)
\thicklines \path(221,179)(241,179)
\thicklines \path(954,179)(934,179)
\put(199,179){\makebox(0,0)[r]{0}}
\thicklines \path(221,332)(241,332)
\thicklines \path(954,332)(934,332)
\put(199,332){\makebox(0,0)[r]{0.04}}
\thicklines \path(221,485)(241,485)
\thicklines \path(954,485)(934,485)
\put(199,485){\makebox(0,0)[r]{0.08}}
\thicklines \path(221,639)(241,639)
\thicklines \path(954,639)(934,639)
\put(199,639){\makebox(0,0)[r]{0.12}}
\thicklines \path(221,792)(241,792)
\thicklines \path(954,792)(934,792)
\put(199,792){\makebox(0,0)[r]{0.16}}
\thicklines \path(221,945)(241,945)
\thicklines \path(954,945)(934,945)
\put(199,945){\makebox(0,0)[r]{0.2}}
\thicklines \path(221,179)(221,199)
\thicklines \path(221,945)(221,925)
\put(221,134){\makebox(0,0){-4}}
\thicklines \path(404,179)(404,199)
\thicklines \path(404,945)(404,925)
\put(404,134){\makebox(0,0){-2}}
\thicklines \path(588,179)(588,199)
\thicklines \path(588,945)(588,925)
\put(588,134){\makebox(0,0){0}}
\thicklines \path(771,179)(771,199)
\thicklines \path(771,945)(771,925)
\put(771,134){\makebox(0,0){2}}
\thicklines \path(954,179)(954,199)
\thicklines \path(954,945)(954,925)
\put(954,134){\makebox(0,0){4}}
\thicklines \path(221,179)(954,179)(954,945)(221,945)(221,179)
\put(45,562){\makebox(0,0)[l]{\shortstack{$D_m$ \\
$\left(\frac{\text{cm}^2}{\text{s}}\right)$}}}
\put(587,67){\makebox(0,0){$z$ (cm)}}
\put(862,868){\makebox(0,0){\large\bf (a)}}

\thinlines \path(290,210)(290,210)(336,209)(381,237)(427,249)(473,239)(519,249)(565,253)(610,242)(656,245)(702,259)(748,244)(794,234)(839,222)(885,203)
\put(290,210){\raisebox{-1.2pt}{\makebox(0,0){$\Diamond$}}}
\put(336,209){\raisebox{-1.2pt}{\makebox(0,0){$\Diamond$}}}
\put(381,237){\raisebox{-1.2pt}{\makebox(0,0){$\Diamond$}}}
\put(427,249){\raisebox{-1.2pt}{\makebox(0,0){$\Diamond$}}}
\put(473,239){\raisebox{-1.2pt}{\makebox(0,0){$\Diamond$}}}
\put(519,249){\raisebox{-1.2pt}{\makebox(0,0){$\Diamond$}}}
\put(565,253){\raisebox{-1.2pt}{\makebox(0,0){$\Diamond$}}}
\put(610,242){\raisebox{-1.2pt}{\makebox(0,0){$\Diamond$}}}
\put(656,245){\raisebox{-1.2pt}{\makebox(0,0){$\Diamond$}}}
\put(702,259){\raisebox{-1.2pt}{\makebox(0,0){$\Diamond$}}}
\put(748,244){\raisebox{-1.2pt}{\makebox(0,0){$\Diamond$}}}
\put(794,234){\raisebox{-1.2pt}{\makebox(0,0){$\Diamond$}}}
\put(839,222){\raisebox{-1.2pt}{\makebox(0,0){$\Diamond$}}}
\put(885,203){\raisebox{-1.2pt}{\makebox(0,0){$\Diamond$}}}
\thinlines \path(290,193)(290,193)(336,221)(381,274)(427,363)(473,474)(519,496)(565,476)(610,469)(656,425)(702,368)(748,317)(794,276)(839,229)(885,205)
\put(290,193){\raisebox{-1.2pt}{\makebox(0,0){$\Box$}}}
\put(336,221){\raisebox{-1.2pt}{\makebox(0,0){$\Box$}}}
\put(381,274){\raisebox{-1.2pt}{\makebox(0,0){$\Box$}}}
\put(427,363){\raisebox{-1.2pt}{\makebox(0,0){$\Box$}}}
\put(473,474){\raisebox{-1.2pt}{\makebox(0,0){$\Box$}}}
\put(519,496){\raisebox{-1.2pt}{\makebox(0,0){$\Box$}}}
\put(565,476){\raisebox{-1.2pt}{\makebox(0,0){$\Box$}}}
\put(610,469){\raisebox{-1.2pt}{\makebox(0,0){$\Box$}}}
\put(656,425){\raisebox{-1.2pt}{\makebox(0,0){$\Box$}}}
\put(702,368){\raisebox{-1.2pt}{\makebox(0,0){$\Box$}}}
\put(748,317){\raisebox{-1.2pt}{\makebox(0,0){$\Box$}}}
\put(794,276){\raisebox{-1.2pt}{\makebox(0,0){$\Box$}}}
\put(839,229){\raisebox{-1.2pt}{\makebox(0,0){$\Box$}}}
\put(885,205){\raisebox{-1.2pt}{\makebox(0,0){$\Box$}}}
\thinlines \path(290,191)(290,191)(336,199)(381,210)(427,217)(473,216)(519,223)(565,227)(610,228)(656,230)(702,222)(748,222)(794,217)(839,210)(885,208)
\put(290,191){\makebox(0,0){$+$}}
\put(336,199){\makebox(0,0){$+$}}
\put(381,210){\makebox(0,0){$+$}}
\put(427,217){\makebox(0,0){$+$}}
\put(473,216){\makebox(0,0){$+$}}
\put(519,223){\makebox(0,0){$+$}}
\put(565,227){\makebox(0,0){$+$}}
\put(610,228){\makebox(0,0){$+$}}
\put(656,230){\makebox(0,0){$+$}}
\put(702,222){\makebox(0,0){$+$}}
\put(748,222){\makebox(0,0){$+$}}
\put(794,217){\makebox(0,0){$+$}}
\put(839,210){\makebox(0,0){$+$}}
\put(885,208){\makebox(0,0){$+$}}
\thinlines \path(290,192)(290,192)(336,208)(381,231)(427,256)(473,297)(519,331)(565,351)(610,518)(656,580)(702,479)(748,412)(794,374)(839,291)(885,237)
\put(290,192){\makebox(0,0){$\times$}}
\put(336,208){\makebox(0,0){$\times$}}
\put(381,231){\makebox(0,0){$\times$}}
\put(427,256){\makebox(0,0){$\times$}}
\put(473,297){\makebox(0,0){$\times$}}
\put(519,331){\makebox(0,0){$\times$}}
\put(565,351){\makebox(0,0){$\times$}}
\put(610,518){\makebox(0,0){$\times$}}
\put(656,580){\makebox(0,0){$\times$}}
\put(702,479){\makebox(0,0){$\times$}}
\put(748,412){\makebox(0,0){$\times$}}
\put(794,374){\makebox(0,0){$\times$}}
\put(839,291){\makebox(0,0){$\times$}}
\put(885,237){\makebox(0,0){$\times$}}
\thinlines \path(290,197)(290,197)(336,223)(381,250)(427,261)(473,316)(519,392)(565,495)(610,620)(656,850)(702,928)(748,736)(794,530)(839,436)(885,234)
\put(290,197){\makebox(0,0){$\triangle$}}
\put(336,223){\makebox(0,0){$\triangle$}}
\put(381,250){\makebox(0,0){$\triangle$}}
\put(427,261){\makebox(0,0){$\triangle$}}
\put(473,316){\makebox(0,0){$\triangle$}}
\put(519,392){\makebox(0,0){$\triangle$}}
\put(565,495){\makebox(0,0){$\triangle$}}
\put(610,620){\makebox(0,0){$\triangle$}}
\put(656,850){\makebox(0,0){$\triangle$}}
\put(702,928){\makebox(0,0){$\triangle$}}
\put(748,736){\makebox(0,0){$\triangle$}}
\put(794,530){\makebox(0,0){$\triangle$}}
\put(839,436){\makebox(0,0){$\triangle$}}
\put(885,234){\makebox(0,0){$\triangle$}}

\path(245,915)(490,915)(490,680)(245,680)(245,915)
\put(343,883){\makebox(0,0)[r]{unary}}
\thinlines \path(365,883)(473,883)
\put(419,883){\raisebox{-1.2pt}{\makebox(0,0){$\Diamond$}}}
\put(343,838){\makebox(0,0)[r]{$0.5$}}
\thinlines \path(365,838)(473,838)
\put(419,838){\raisebox{-1.2pt}{\makebox(0,0){$\Box$}}}
\put(343,793){\makebox(0,0)[r]{$1.0$}}
\thinlines \path(365,793)(473,793)
\put(419,793){\makebox(0,0){$+$}}
\put(343,748){\makebox(0,0)[r]{$2.0$}}
\thinlines \path(365,748)(473,748)
\put(419,748){\makebox(0,0){$\times$}}
\put(343,703){\makebox(0,0)[r]{$4.0$}}
\thinlines \path(365,703)(473,703)
\put(419,703){\makebox(0,0){$\triangle$}}

\end{picture} \hspace{-3.6cm} 
\setlength{\unitlength}{0.240900pt}
\begin{picture}(1500,920)(0,80)
\thicklines \path(221,179)(241,179)
\thicklines \path(954,179)(934,179)
\put(199,179){\makebox(0,0)[r]{0}}
\thicklines \path(221,332)(241,332)
\thicklines \path(954,332)(934,332)
\put(199,332){\makebox(0,0)[r]{0.02}}
\thicklines \path(221,485)(241,485)
\thicklines \path(954,485)(934,485)
\put(199,485){\makebox(0,0)[r]{0.04}}
\thicklines \path(221,639)(241,639)
\thicklines \path(954,639)(934,639)
\put(199,639){\makebox(0,0)[r]{0.06}}
\thicklines \path(221,792)(241,792)
\thicklines \path(954,792)(934,792)
\put(199,792){\makebox(0,0)[r]{0.08}}
\thicklines \path(221,945)(241,945)
\thicklines \path(954,945)(934,945)
\put(199,945){\makebox(0,0)[r]{0.1}}
\thicklines \path(221,179)(221,199)
\thicklines \path(221,945)(221,925)
\put(221,134){\makebox(0,0){-4}}
\thicklines \path(404,179)(404,199)
\thicklines \path(404,945)(404,925)
\put(404,134){\makebox(0,0){-2}}
\thicklines \path(588,179)(588,199)
\thicklines \path(588,945)(588,925)
\put(588,134){\makebox(0,0){0}}
\thicklines \path(771,179)(771,199)
\thicklines \path(771,945)(771,925)
\put(771,134){\makebox(0,0){2}}
\thicklines \path(954,179)(954,199)
\thicklines \path(954,945)(954,925)
\put(954,134){\makebox(0,0){4}}
\thicklines \path(221,179)(954,179)(954,945)(221,945)(221,179)
\put(45,562){\makebox(0,0)[l]{\shortstack{$D_m$ \\
$\left(\frac{\text{cm}^2}{\text{s}}\right)$}}}
\put(587,67){\makebox(0,0){z (cm)}}
\put(312,868){\makebox(0,0){\large\bf (b)}}

\thinlines \path(290,207)(290,207)(336,264)(381,369)(427,547)(473,769)(519,812)(565,773)(610,758)(656,670)(702,556)(748,455)(794,373)(839,280)(885,231)
\put(290,207){\raisebox{-1.2pt}{\makebox(0,0){$\Box$}}}
\put(336,264){\raisebox{-1.2pt}{\makebox(0,0){$\Box$}}}
\put(381,369){\raisebox{-1.2pt}{\makebox(0,0){$\Box$}}}
\put(427,547){\raisebox{-1.2pt}{\makebox(0,0){$\Box$}}}
\put(473,769){\raisebox{-1.2pt}{\makebox(0,0){$\Box$}}}
\put(519,812){\raisebox{-1.2pt}{\makebox(0,0){$\Box$}}}
\put(565,773){\raisebox{-1.2pt}{\makebox(0,0){$\Box$}}}
\put(610,758){\raisebox{-1.2pt}{\makebox(0,0){$\Box$}}}
\put(656,670){\raisebox{-1.2pt}{\makebox(0,0){$\Box$}}}
\put(702,556){\raisebox{-1.2pt}{\makebox(0,0){$\Box$}}}
\put(748,455){\raisebox{-1.2pt}{\makebox(0,0){$\Box$}}}
\put(794,373){\raisebox{-1.2pt}{\makebox(0,0){$\Box$}}}
\put(839,280){\raisebox{-1.2pt}{\makebox(0,0){$\Box$}}}
\put(885,231){\raisebox{-1.2pt}{\makebox(0,0){$\Box$}}}
\thinlines \path(290,207)(290,207)(336,264)(381,369)(427,547)(473,769)(519,812)(565,736)(610,633)(656,460)(702,210)
\put(290,207){\makebox(0,0){$\times$}}
\put(336,264){\makebox(0,0){$\times$}}
\put(381,369){\makebox(0,0){$\times$}}
\put(427,547){\makebox(0,0){$\times$}}
\put(473,769){\makebox(0,0){$\times$}}
\put(519,812){\makebox(0,0){$\times$}}
\put(565,736){\makebox(0,0){$\times$}}
\put(610,633){\makebox(0,0){$\times$}}
\put(656,460){\makebox(0,0){$\times$}}
\put(702,210){\makebox(0,0){$\times$}}
\thinlines \path(519,293)(565,652)(610,716)(656,673)(702,558)(748,455)(794,373)(839,280)(885,231)
\put(519,293){\makebox(0,0){$\triangle$}}
\put(565,652){\makebox(0,0){$\triangle$}}
\put(610,716){\makebox(0,0){$\triangle$}}
\put(656,673){\makebox(0,0){$\triangle$}}
\put(702,558){\makebox(0,0){$\triangle$}}
\put(748,455){\makebox(0,0){$\triangle$}}
\put(794,373){\makebox(0,0){$\triangle$}}
\put(839,280){\makebox(0,0){$\triangle$}}
\put(885,231){\makebox(0,0){$\triangle$}}

\path(680,912)(920,912)(920,762)(680,762)(680,912)
\put(780,882){\makebox(0,0)[r]{both}}
\thinlines \path(802,882)(910,882)
\put(856,882){\raisebox{-1.2pt}{\makebox(0,0){$\Box$}}}
\put(780,837){\makebox(0,0)[r]{small}}
\thinlines \path(802,837)(910,837)
\put(856,837){\makebox(0,0){$\times$}}
\put(780,792){\makebox(0,0)[r]{large}}
\thinlines \path(802,792)(910,792)
\put(856,792){\makebox(0,0){$\triangle$}}

\end{picture} }
  \caption{Microscopic calculated diffusion coefficients: (a) for different
           density ratios $\rho/\rho_l$ and (b) calculated separately for the 
	   small and large particles for $\rho/\rho_l=0.5$.}
  \label{fig: micro_1}
\end{figure*}
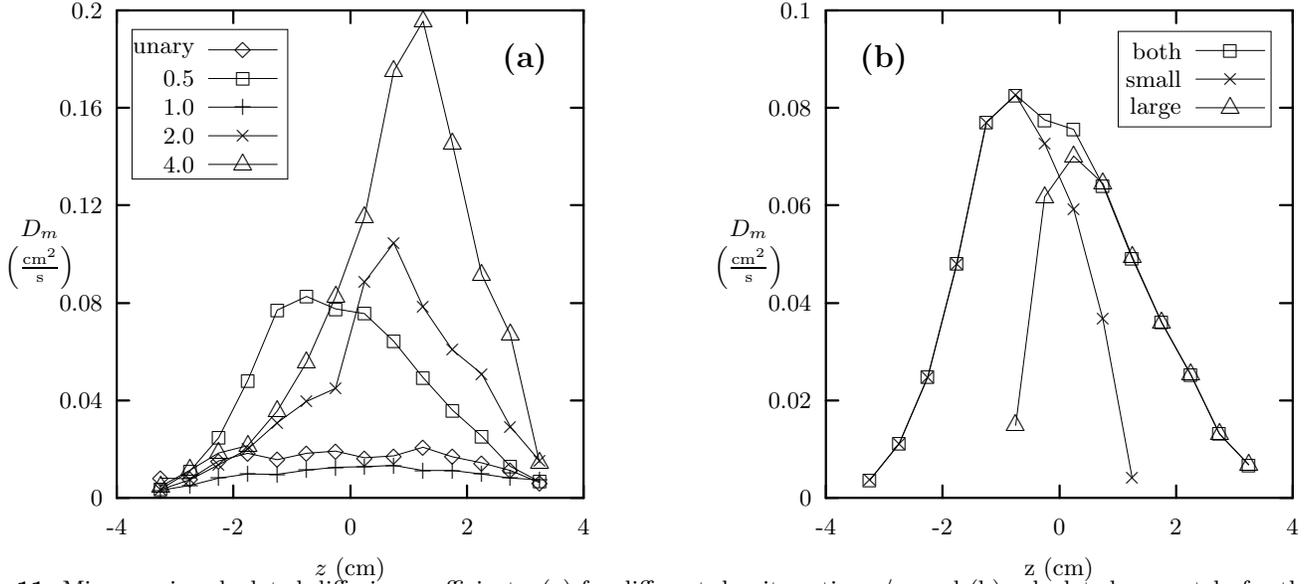

The initial particle rearrangement when the cylinder starts to rotate cannot be
described by a diffusion process. Therefore, we begin all our measurements at
the point when the continuous flow has set in which corresponds to one eighth
of a  cylinder revolution. In order to resolve the diffusion process spatially,
we divide the cylinder into 14 equal slices along the rotational axis and
calculate the microscopic diffusion coefficient in each slice from the particle
trajectories that start in the corresponding slice. This is done for different
density ratios, $\rho/\rho_l$, of a binary particle mixture and shown in
Fig.~\ref{fig: micro_1}a. Changing the density ratio has a dramatic effect on
the diffusion coefficient, increasing the maximum of $D_m$ by a factor of ten
when the density of the smaller particles is increased by a factor of 4. In the
case where $\rho/\rho_l > 1$, the maximum diffusion coefficient can be found in
the region of the larger particles ($z>0$ cm) and for $\rho/\rho_l < 1$ in the
region of the smaller particles; i.e. the spatial maximum of $D_m$ lies on the
side of the lighter particles for all density ratios. Nevertheless, in each
case the maximum value of the diffusion coefficient is close to the middle.

For comparison, we also calculated the diffusion coefficient for a system with
only the larger particle component which is referred to as {\em unary} mixture.
One thing to note is that the diffusion coefficients for the binary mixture
with equal density, denoted by a cross ($+$) and the unary  mixture, denoted by
a diamond ($\Diamond$) in the following plots, nearly agree despite the size
difference in the binary mixture and hardly show a spatial variation.

In Fig.~\ref{fig: micro_1}b, we calculate $D_m$ separately for the smaller and
larger particles using a density ratio of $\rho/\rho_l=0.5$. The squares
($\Box$) denote the average diffusion coefficient, as already shown in
Fig.~\ref{fig: micro_1}a and the crosses ($\times$) and triangles ($\triangle$)
stand for the  diffusion coefficient of the smaller and larger particles,
respectively. In this case, the maximum of $D_m$ for the smaller particles is 
larger than for the larger particles. It is also seen that the maximum
diffusion coefficient for the smaller particles lies in the region where the
small particles have been initially. For the larger particles, the maximum
value of the diffusion coefficient lies in the region where the larger particles
have been initially and both maxima are very close to the initial interface.

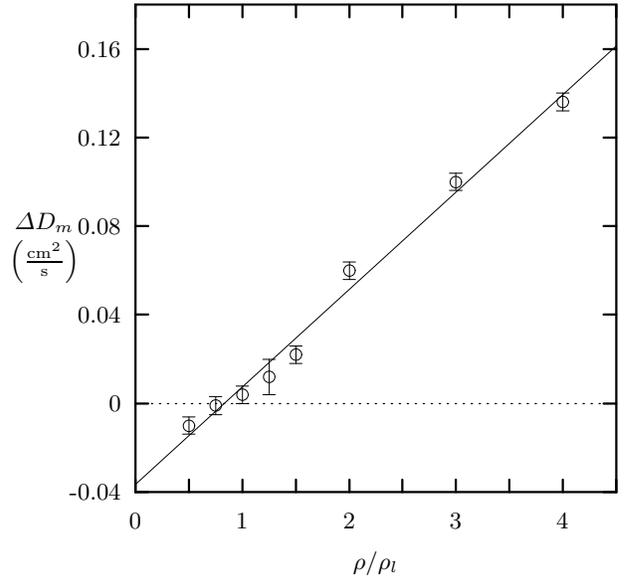
\begin{figure}
  {\centering{
\setlength{\unitlength}{0.240900pt}
\begin{picture}(1500,900)(0,80)
\thicklines \path(243,179)(263,179)
\thicklines \path(998,179)(978,179)
\put(221,179){\makebox(0,0)[r]{-0.04}}
\thicklines \path(243,318)(263,318)
\thicklines \path(998,318)(978,318)
\put(221,318){\makebox(0,0)[r]{0}}
\thicklines \path(243,458)(263,458)
\thicklines \path(998,458)(978,458)
\put(221,458){\makebox(0,0)[r]{0.04}}
\thicklines \path(243,597)(263,597)
\thicklines \path(998,597)(978,597)
\put(221,597){\makebox(0,0)[r]{0.08}}
\thicklines \path(243,736)(263,736)
\thicklines \path(998,736)(978,736)
\put(221,736){\makebox(0,0)[r]{0.12}}
\thicklines \path(243,875)(263,875)
\thicklines \path(998,875)(978,875)
\put(221,875){\makebox(0,0)[r]{0.16}}
\thicklines \path(243,179)(243,199)
\thicklines \path(243,945)(243,925)
\put(243,134){\makebox(0,0){0}}
\thicklines \path(327,179)(327,199)
\thicklines \path(327,945)(327,925)
\put(327,134){\makebox(0,0){}}
\thicklines \path(411,179)(411,199)
\thicklines \path(411,945)(411,925)
\put(411,134){\makebox(0,0){1}}
\thicklines \path(495,179)(495,199)
\thicklines \path(495,945)(495,925)
\put(495,134){\makebox(0,0){}}
\thicklines \path(579,179)(579,199)
\thicklines \path(579,945)(579,925)
\put(579,134){\makebox(0,0){2}}
\thicklines \path(662,179)(662,199)
\thicklines \path(662,945)(662,925)
\put(662,134){\makebox(0,0){}}
\thicklines \path(746,179)(746,199)
\thicklines \path(746,945)(746,925)
\put(746,134){\makebox(0,0){3}}
\thicklines \path(830,179)(830,199)
\thicklines \path(830,945)(830,925)
\put(830,134){\makebox(0,0){}}
\thicklines \path(914,179)(914,199)
\thicklines \path(914,945)(914,925)
\put(914,134){\makebox(0,0){4}}
\thicklines \path(998,179)(998,199)
\thicklines \path(998,945)(998,925)
\put(998,134){\makebox(0,0){}}
\thicklines \path(243,179)(998,179)(998,945)(243,945)(243,179)
\put(45,562){\makebox(0,0)[l]{\shortstack{$\Delta D_m$ \\ $\left(\frac{\text{cm}^2}{\text{s}}\right)$}}}
\put(620,67){\makebox(0,0){$\rho/\rho_l$}}
\thinlines \path(327,270)(327,297)
\thinlines \path(317,270)(337,270)
\thinlines \path(317,297)(337,297)
\thinlines \path(369,301)(369,329)
\thinlines \path(359,301)(379,301)
\thinlines \path(359,329)(379,329)
\thinlines \path(411,318)(411,346)
\thinlines \path(401,318)(421,318)
\thinlines \path(401,346)(421,346)
\thinlines \path(453,332)(453,388)
\thinlines \path(443,332)(463,332)
\thinlines \path(443,388)(463,388)
\thinlines \path(495,381)(495,409)
\thinlines \path(485,381)(505,381)
\thinlines \path(485,409)(505,409)
\thinlines \path(579,513)(579,541)
\thinlines \path(569,513)(589,513)
\thinlines \path(569,541)(589,541)
\thinlines \path(746,653)(746,680)
\thinlines \path(736,653)(756,653)
\thinlines \path(736,680)(756,680)
\thinlines \path(914,778)(914,806)
\thinlines \path(904,778)(924,778)
\thinlines \path(904,806)(924,806)
\put(327,283){\circle{18}}
\put(369,315){\circle{18}}
\put(411,332){\circle{18}}
\put(453,360){\circle{18}}
\put(495,395){\circle{18}}
\put(579,527){\circle{18}}
\put(746,666){\circle{18}}
\put(914,792){\circle{18}}
\path(243,191)(998,880)
\dottedline{13}(243,318)(998,318)
\end{picture} }}
  \caption{Difference in the maximum value of the diffusion coefficient of the 
           larger and smaller particles as function of density ratio. The linear
	   least-square fit is shown as solid line.}
\label{fig: delta_d}
\end{figure}

The difference in the maximum value of the diffusion coefficient of the larger
and the smaller particles, $\Delta D_m$, is shown in Fig.~\ref{fig: delta_d} as
function of the density ratio. For a value of $\rho/\rho_l\gtrsim 1$, the
larger particles have a larger maximum diffusion coefficient which agrees with
the experimental observation that for particles with the same density the large
particles have a higher mobility than the smaller particles~\cite{dasgupta91}.
For values $\rho/\rho_l<1$, the mobility of the smaller particles is higher,
resulting in a negative difference in Fig.~\ref{fig: delta_d}. The linear
least-square fit using all data points is shown as solid line.

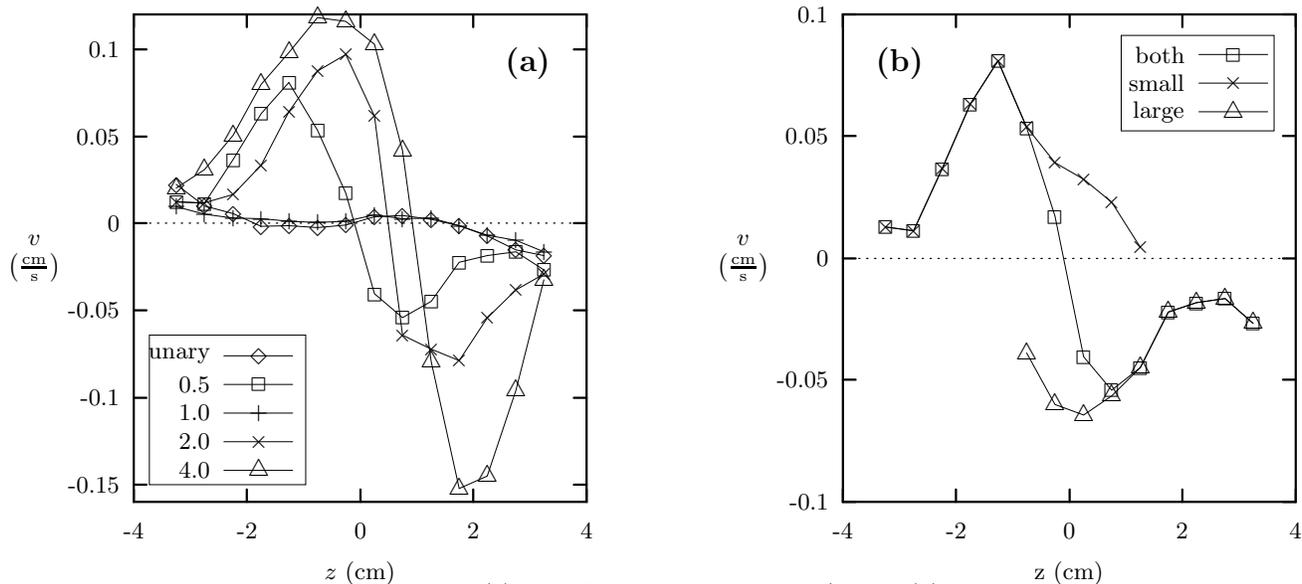
\begin{figure*}
  \hbox{ 
\setlength{\unitlength}{0.240900pt}
\begin{picture}(1500,920)(0,80)
\thicklines \path(243,206)(263,206)
\thicklines \path(954,206)(934,206)
\put(221,206){\makebox(0,0)[r]{-0.15}}
\thicklines \path(243,343)(263,343)
\thicklines \path(954,343)(934,343)
\put(221,343){\makebox(0,0)[r]{-0.1}}
\thicklines \path(243,480)(263,480)
\thicklines \path(954,480)(934,480)
\put(221,480){\makebox(0,0)[r]{-0.05}}
\thicklines \path(243,617)(263,617)
\thicklines \path(954,617)(934,617)
\put(221,617){\makebox(0,0)[r]{0}}
\thicklines \path(243,753)(263,753)
\thicklines \path(954,753)(934,753)
\put(221,753){\makebox(0,0)[r]{0.05}}
\thicklines \path(243,890)(263,890)
\thicklines \path(954,890)(934,890)
\put(221,890){\makebox(0,0)[r]{0.1}}
\thicklines \path(243,179)(243,199)
\thicklines \path(243,945)(243,925)
\put(243,134){\makebox(0,0){-4}}
\thicklines \path(421,179)(421,199)
\thicklines \path(421,945)(421,925)
\put(421,134){\makebox(0,0){-2}}
\thicklines \path(599,179)(599,199)
\thicklines \path(599,945)(599,925)
\put(599,134){\makebox(0,0){0}}
\thicklines \path(776,179)(776,199)
\thicklines \path(776,945)(776,925)
\put(776,134){\makebox(0,0){2}}
\thicklines \path(954,179)(954,199)
\thicklines \path(954,945)(954,925)
\put(954,134){\makebox(0,0){4}}
\thicklines \path(243,179)(954,179)(954,945)(243,945)(243,179)
\put(45,562){\makebox(0,0)[l]{\shortstack{$v$ \\
$\left(\frac{\text{cm}}{\text{s}}\right)$}}}
\put(598,67){\makebox(0,0){$z$ (cm)}}
\put(862,868){\makebox(0,0){\large\bf (a)}}

\thinlines \path(310,678)(310,678)(354,644)(399,632)(443,612)(487,613)(532,610)(576,614)(621,627)(665,629)(710,623)(754,613)(798,597)(843,575)(887,566)
\put(310,678){\raisebox{-1.2pt}{\makebox(0,0){$\Diamond$}}}
\put(354,644){\raisebox{-1.2pt}{\makebox(0,0){$\Diamond$}}}
\put(399,632){\raisebox{-1.2pt}{\makebox(0,0){$\Diamond$}}}
\put(443,612){\raisebox{-1.2pt}{\makebox(0,0){$\Diamond$}}}
\put(487,613){\raisebox{-1.2pt}{\makebox(0,0){$\Diamond$}}}
\put(532,610){\raisebox{-1.2pt}{\makebox(0,0){$\Diamond$}}}
\put(576,614){\raisebox{-1.2pt}{\makebox(0,0){$\Diamond$}}}
\put(621,627){\raisebox{-1.2pt}{\makebox(0,0){$\Diamond$}}}
\put(665,629){\raisebox{-1.2pt}{\makebox(0,0){$\Diamond$}}}
\put(710,623){\raisebox{-1.2pt}{\makebox(0,0){$\Diamond$}}}
\put(754,613){\raisebox{-1.2pt}{\makebox(0,0){$\Diamond$}}}
\put(798,597){\raisebox{-1.2pt}{\makebox(0,0){$\Diamond$}}}
\put(843,575){\raisebox{-1.2pt}{\makebox(0,0){$\Diamond$}}}
\put(887,566){\raisebox{-1.2pt}{\makebox(0,0){$\Diamond$}}}
\thinlines \path(310,651)(310,651)(354,648)(399,717)(443,789)(487,838)(532,763)(576,664)(621,506)(665,469)(710,494)(754,556)(798,566)(843,572)(887,544)
\put(310,651){\raisebox{-1.2pt}{\makebox(0,0){$\Box$}}}
\put(354,648){\raisebox{-1.2pt}{\makebox(0,0){$\Box$}}}
\put(399,717){\raisebox{-1.2pt}{\makebox(0,0){$\Box$}}}
\put(443,789){\raisebox{-1.2pt}{\makebox(0,0){$\Box$}}}
\put(487,838){\raisebox{-1.2pt}{\makebox(0,0){$\Box$}}}
\put(532,763){\raisebox{-1.2pt}{\makebox(0,0){$\Box$}}}
\put(576,664){\raisebox{-1.2pt}{\makebox(0,0){$\Box$}}}
\put(621,506){\raisebox{-1.2pt}{\makebox(0,0){$\Box$}}}
\put(665,469){\raisebox{-1.2pt}{\makebox(0,0){$\Box$}}}
\put(710,494){\raisebox{-1.2pt}{\makebox(0,0){$\Box$}}}
\put(754,556){\raisebox{-1.2pt}{\makebox(0,0){$\Box$}}}
\put(798,566){\raisebox{-1.2pt}{\makebox(0,0){$\Box$}}}
\put(843,572){\raisebox{-1.2pt}{\makebox(0,0){$\Box$}}}
\put(887,544){\raisebox{-1.2pt}{\makebox(0,0){$\Box$}}}
\thinlines \path(310,643)(310,643)(354,631)(399,625)(443,624)(487,620)(532,618)(576,620)(621,630)(665,624)(710,625)(754,613)(798,598)(843,590)(887,572)
\put(310,643){\makebox(0,0){$+$}}
\put(354,631){\makebox(0,0){$+$}}
\put(399,625){\makebox(0,0){$+$}}
\put(443,624){\makebox(0,0){$+$}}
\put(487,620){\makebox(0,0){$+$}}
\put(532,618){\makebox(0,0){$+$}}
\put(576,620){\makebox(0,0){$+$}}
\put(621,630){\makebox(0,0){$+$}}
\put(665,624){\makebox(0,0){$+$}}
\put(710,625){\makebox(0,0){$+$}}
\put(754,613){\makebox(0,0){$+$}}
\put(798,598){\makebox(0,0){$+$}}
\put(843,590){\makebox(0,0){$+$}}
\put(887,572){\makebox(0,0){$+$}}
\thinlines \path(310,650)(310,650)(354,649)(399,662)(443,708)(487,792)(532,856)(576,882)(621,785)(665,441)(710,419)(754,401)(798,468)(843,512)(887,537)
\put(310,650){\makebox(0,0){$\times$}}
\put(354,649){\makebox(0,0){$\times$}}
\put(399,662){\makebox(0,0){$\times$}}
\put(443,708){\makebox(0,0){$\times$}}
\put(487,792){\makebox(0,0){$\times$}}
\put(532,856){\makebox(0,0){$\times$}}
\put(576,882){\makebox(0,0){$\times$}}
\put(621,785){\makebox(0,0){$\times$}}
\put(665,441){\makebox(0,0){$\times$}}
\put(710,419){\makebox(0,0){$\times$}}
\put(754,401){\makebox(0,0){$\times$}}
\put(798,468){\makebox(0,0){$\times$}}
\put(843,512){\makebox(0,0){$\times$}}
\put(887,537){\makebox(0,0){$\times$}}
\thinlines \path(310,672)(310,672)(354,701)(399,755)(443,835)(487,885)(532,940)(576,934)(621,898)(665,731)(710,400)(754,200)(798,220)(843,354)(887,528)
\put(310,672){\makebox(0,0){$\triangle$}}
\put(354,701){\makebox(0,0){$\triangle$}}
\put(399,755){\makebox(0,0){$\triangle$}}
\put(443,835){\makebox(0,0){$\triangle$}}
\put(487,885){\makebox(0,0){$\triangle$}}
\put(532,940){\makebox(0,0){$\triangle$}}
\put(576,934){\makebox(0,0){$\triangle$}}
\put(621,898){\makebox(0,0){$\triangle$}}
\put(665,731){\makebox(0,0){$\triangle$}}
\put(710,400){\makebox(0,0){$\triangle$}}
\put(754,200){\makebox(0,0){$\triangle$}}
\put(798,220){\makebox(0,0){$\triangle$}}
\put(843,354){\makebox(0,0){$\triangle$}}
\put(887,528){\makebox(0,0){$\triangle$}}
\dottedline{13}(243,617)(954,617)

\path(268,441)(513,441)(513,206)(268,206)(268,441)
\put(363,409){\makebox(0,0)[r]{unary}}
\thinlines \path(385,409)(493,409)
\put(439,409){\raisebox{-1.2pt}{\makebox(0,0){$\Diamond$}}}
\put(363,364){\makebox(0,0)[r]{$0.5$}}
\thinlines \path(385,364)(493,364)
\put(439,364){\raisebox{-1.2pt}{\makebox(0,0){$\Box$}}}
\put(363,319){\makebox(0,0)[r]{$1.0$}}
\thinlines \path(385,319)(493,319)
\put(439,319){\makebox(0,0){$+$}}
\put(363,274){\makebox(0,0)[r]{$2.0$}}
\thinlines \path(385,274)(493,274)
\put(439,274){\makebox(0,0){$\times$}}
\put(363,229){\makebox(0,0)[r]{$4.0$}}
\thinlines \path(385,229)(493,229)
\put(439,229){\makebox(0,0){$\triangle$}}

\end{picture} \hspace{-3.6cm} 
\setlength{\unitlength}{0.240900pt}
\begin{picture}(1500,920)(0,80)
\thicklines \path(243,179)(263,179)
\thicklines \path(954,179)(934,179)
\put(221,179){\makebox(0,0)[r]{-0.1}}
\thicklines \path(243,371)(263,371)
\thicklines \path(954,371)(934,371)
\put(221,371){\makebox(0,0)[r]{-0.05}}
\thicklines \path(243,562)(263,562)
\thicklines \path(954,562)(934,562)
\put(221,562){\makebox(0,0)[r]{0}}
\thicklines \path(243,754)(263,754)
\thicklines \path(954,754)(934,754)
\put(221,754){\makebox(0,0)[r]{0.05}}
\thicklines \path(243,945)(263,945)
\thicklines \path(954,945)(934,945)
\put(221,945){\makebox(0,0)[r]{0.1}}
\thicklines \path(243,179)(243,199)
\thicklines \path(243,945)(243,925)
\put(243,134){\makebox(0,0){-4}}
\thicklines \path(421,179)(421,199)
\thicklines \path(421,945)(421,925)
\put(421,134){\makebox(0,0){-2}}
\thicklines \path(599,179)(599,199)
\thicklines \path(599,945)(599,925)
\put(599,134){\makebox(0,0){0}}
\thicklines \path(776,179)(776,199)
\thicklines \path(776,945)(776,925)
\put(776,134){\makebox(0,0){2}}
\thicklines \path(954,179)(954,199)
\thicklines \path(954,945)(954,925)
\put(954,134){\makebox(0,0){4}}
\thicklines \path(243,179)(954,179)(954,945)(243,945)(243,179)
\put(45,562){\makebox(0,0)[l]{\shortstack{$v$ \\
$\left(\frac{\text{cm}}{\text{s}}\right)$
}}}
\put(598,67){\makebox(0,0){z (cm)}}
\put(332,868){\makebox(0,0){\large\bf (b)}}

\thinlines \path(310,611)(310,611)(354,605)(399,703)(443,804)(487,872)(532,767)(576,628)(621,407)(665,355)(710,390)(754,477)(798,492)(843,499)(887,460)
\put(310,611){\raisebox{-1.2pt}{\makebox(0,0){$\Box$}}}
\put(354,605){\raisebox{-1.2pt}{\makebox(0,0){$\Box$}}}
\put(399,703){\raisebox{-1.2pt}{\makebox(0,0){$\Box$}}}
\put(443,804){\raisebox{-1.2pt}{\makebox(0,0){$\Box$}}}
\put(487,872){\raisebox{-1.2pt}{\makebox(0,0){$\Box$}}}
\put(532,767){\raisebox{-1.2pt}{\makebox(0,0){$\Box$}}}
\put(576,628){\raisebox{-1.2pt}{\makebox(0,0){$\Box$}}}
\put(621,407){\raisebox{-1.2pt}{\makebox(0,0){$\Box$}}}
\put(665,355){\raisebox{-1.2pt}{\makebox(0,0){$\Box$}}}
\put(710,390){\raisebox{-1.2pt}{\makebox(0,0){$\Box$}}}
\put(754,477){\raisebox{-1.2pt}{\makebox(0,0){$\Box$}}}
\put(798,492){\raisebox{-1.2pt}{\makebox(0,0){$\Box$}}}
\put(843,499){\raisebox{-1.2pt}{\makebox(0,0){$\Box$}}}
\put(887,460){\raisebox{-1.2pt}{\makebox(0,0){$\Box$}}}
\thinlines \path(310,611)(310,611)(354,605)(399,703)(443,804)(487,872)(532,768)(576,712)(621,686)(665,650)(710,579)
\put(310,611){\makebox(0,0){$\times$}}
\put(354,605){\makebox(0,0){$\times$}}
\put(399,703){\makebox(0,0){$\times$}}
\put(443,804){\makebox(0,0){$\times$}}
\put(487,872){\makebox(0,0){$\times$}}
\put(532,768){\makebox(0,0){$\times$}}
\put(576,712){\makebox(0,0){$\times$}}
\put(621,686){\makebox(0,0){$\times$}}
\put(665,650){\makebox(0,0){$\times$}}
\put(710,579){\makebox(0,0){$\times$}}
\thinlines \path(532,413)(576,332)(621,316)(665,346)(710,390)(754,478)(798,492)(843,499)(887,460)
\put(532,413){\makebox(0,0){$\triangle$}}
\put(576,332){\makebox(0,0){$\triangle$}}
\put(621,316){\makebox(0,0){$\triangle$}}
\put(665,346){\makebox(0,0){$\triangle$}}
\put(710,390){\makebox(0,0){$\triangle$}}
\put(754,478){\makebox(0,0){$\triangle$}}
\put(798,492){\makebox(0,0){$\triangle$}}
\put(843,499){\makebox(0,0){$\triangle$}}
\put(887,460){\makebox(0,0){$\triangle$}}
\dottedline{13}(243,562)(954,562)

\path(680,912)(920,912)(920,762)(680,762)(680,912)
\put(780,882){\makebox(0,0)[r]{both}}
\thinlines \path(802,882)(910,882)
\put(856,882){\raisebox{-1.2pt}{\makebox(0,0){$\Box$}}}
\put(780,837){\makebox(0,0)[r]{small}}
\thinlines \path(802,837)(910,837)
\put(856,837){\makebox(0,0){$\times$}}
\put(780,792){\makebox(0,0)[r]{large}}
\thinlines \path(802,792)(910,792)
\put(856,792){\makebox(0,0){$\triangle$}}

\end{picture} }
  \caption{Microscopic calculated drift velocities: (a) for different density
           ratios $\rho/\rho_l$ and (b) calculated separately for the small and 
	   large particles for $\rho/\rho_l=0.5$.}
  \label{fig: micro_3}
\end{figure*} 
\subsection{Drift Velocity}
\label{sec: drift}
By using the definition given in Eq.~(\ref{eq: drift}), we can calculate in
each of the 14 slices an average drift velocity of all particles. This is
plotted in Fig.~\ref{fig: micro_3}a for different density ratios $\rho/\rho_l$.
For comparison, the spatial dependence for a unary mixture is also shown,
($\Diamond$), and as before is very close to the dependence of an equal density
binary mixture, ($+$). A much larger drift velocity with a well pronounced
maximum at $z<0$ and minimum at $z>0$ is observed for $\rho/\rho_l \neq 1$,
showing that the global motion exchanges particles across the interface. The
larger the density ratio, the larger the region with a positive drift velocity.
Consequently, the position corresponding to $v=0$ will move to the right for
{\em increasing} density ratios and move to the left for {\em decreasing}
density ratios. Please note that this description only applies to the situation
shortly after the start of the rotation since a symmetric profile is expected
in the steady state due to the symmetry of the problem. The drift velocity can
be explained by applying a ``hydrostatic picture'' again: The hydrostatic
pressure at the interface is given by Eq.(\ref{eq: pressure}). If we now have
two different values for $\rho$ at the interface, there will be a pressure
difference of 
\begin{equation}
  \Delta p \propto |\rho-\rho_l| g \left(1-e^{\frac{-h_d}{c}}\right)
\end{equation}
which causes the drift. This drift will not happen on the free surface (where
the pressure difference is zero), instead the denser particle will push their
way through the lighter ones near the center of revolution, which is well below
the rotational axis. Even though the ``roller coaster'' effect still applies 
here due to the motion inside the granular material (see section~\ref{sec:
friction} for more details), we get a drift in the case of two different
densities for the particles.

In Fig.~\ref{fig: micro_3}b, the individual drift velocities for the smaller and
larger particles are shown along the rotational axis. The density ratio was
$\rho/\rho_l=0.5$ and one observes that the particles close to the end caps of
the cylinder hardly drift at all. On the other hand, the drift velocity is very
large for both components in regions close to the initial interface. Since the
initial stage of the interface dynamics was investigated, the drift velocity is
positive everywhere for the smaller particles and negative everywhere for the
larger particles which clearly shows the particle exchange over the initial
interface.

\section{Conclusions}
We started with an initially sharp front of particles with different properties
and looked how this front gets diffused. In the first part of this paper we
showed that this process can be well approximated by a pure diffusional
process, which was originally applied only to the case of one particle
component.  Also we found that radial segregation hinders diffusion, because
the process of radial segregation will sort the smaller particles out which are
then unable to take place in the diffusion process.

When changing the density of the small particles, we get least mixing for
particles with  the same density. For different density there is a pressure
difference at the interface and the denser particles will penetrate into the
lighter ones. This is not a pure diffusional process anymore, because of the
non-vanishing drift of the particles, it is more like a core movement combined
with diffusion.

Previous to this work the core was mostly thought to be a solid block in which
no (or just minimal) movements can take place. We now showed that for different
particle densities, and maybe also for differences in other particle
properties, core movement is indeed possible.  This could shed light on future
work on the axial segregation mechanism.

\section*{Acknowledgments}
We would like to thank the HLRZ in J\"ulich and the HRZ Marburg for supporting
us with a generous grant of computer time on their Cray T3E and IBM SP2,
respectively. Financial support by the Deutsche Forschungsgemeinschaft is also
greatfully acknowledged.


\end{document}